\documentclass[
preprint,
amsmath,
amssymb,
aps,
tightenlines,
]{revtex4-2}

\usepackage{amsmath}
\usepackage{amssymb}
\usepackage{graphicx}%
\usepackage{dcolumn}%
\usepackage{bm}%
\usepackage{soul} %
\usepackage{xcolor} %

\appdef \turnpage {%
  \AddToHookNext{shipout/after}{%
    \global\pdfpageattr\expandafter{\the\pdfpageattr/Rotate 90}%
    \AddToHookNext{shipout/after}{%
      \global\pdfpageattr\expandafter{\the\pdfpageattr/Rotate 0}%
    }%
  }%
}

\newcommand{\reals}{\mathbb{R}}
\newcommand{\md}{\mathrm{d}}
\newcommand{\grad}{\nabla}

\newcommand{\thermb}{\beta}
\newcommand{\kbt}{\mathrm{k_B}T}

\newcommand{\fgr}{\mathbf{r}}
\newcommand{\fgn}{n}
\newcommand{\fgff}{u}
\newcommand{\fgforce}{f}

\newcommand{\cgr}{\mathbf{R}}

\newcommand{\cgn}{N}
\newcommand{\cgff}{U}
\newcommand{\cgforce}{F}
\newcommand{\cgffparam}{\theta}

\newcommand{\mbpmf}{\cgff_\mathrm{PMF}}
\newcommand{\cmap}{\mathbf{M}}

\newcommand{\kernel}{\kappa}

\newcommand{\forceresidual}{\mathcal{L}}
\newcommand{\fmap}{\mathbf{T}}

\newcommand{\raverage}[1]{
    \left \langle {#1} \right \rangle
}

\newcommand{\density}{p}
\newcommand{\covmat}{\Sigma}
\newcommand{\expect}{\mathbb{E}}

\newcommand{\ddmdensity}{\kernel}
\newcommand{\ddmr}{\fgr}
\newcommand{\ddmR}{\fgr'}

\newcommand{\scorecandidate}{S}

\begin{document}

\title{Learning data efficient coarse-grained molecular dynamics from forces and noise}

\author{Aleksander E. P. Durumeric$^1$}
\affiliation{Department of Mathematics and Computer Science, Freie Universit\"{a}t Berlin, Arnimallee 12, 14195 Berlin, Germany}
\author{Yaoyi Chen$^1$}
\affiliation{Department of Mathematics and Computer Science, Freie Universit\"{a}t Berlin, Arnimallee 12, 14195 Berlin, Germany}
\author{Frank No\'{e}$^*$}
\affiliation{Department of Mathematics and Computer Science, Freie Universit\"{a}t Berlin, Arnimallee 12, 14195 Berlin, Germany}
\affiliation{Department of Physics, Freie Universit\"{a}t Berlin, Arnimallee 12, 14195 Berlin, Germany}
\affiliation{AI for Science, Microsoft Research, Karl-Liebknecht Str. 32, 10178 Berlin, Germany}
\author{Cecilia Clementi$^*$}
\affiliation{Department of Physics, Freie Universit\"{a}t Berlin, Arnimallee 12, 14195 Berlin, Germany}
\affiliation{Center for Theoretical Biological Physics, Rice University, Houston,
77005, TX, USA}
\affiliation{Department of Chemistry, Rice University, Houston, 77005, TX, USA.}


\begin{abstract}
Machine-learned coarse-grained (MLCG) molecular dynamics is a promising option for modeling biomolecules. However, MLCG models currently require large amounts of data from reference atomistic molecular dynamics or substantial computation for training. Denoising score matching --- the technology behind the widely popular diffusion models --- has simultaneously emerged as a machine-learning framework for creating samples from noise. Models in the first category are often trained using atomistic forces, 
while those in the second category extract the data distribution by reverting noise-based corruption.
We unify these approaches to improve the training of MLCG force-fields, reducing data requirements by a factor of 100 while maintaining advantages typical to force-based parameterization.
The methods are demonstrated on proteins Trp-Cage and NTL9 and published as open-source code.
\end{abstract}

\maketitle

Atomistic molecular dynamics (MD) can, in principle, connect the microscopic motion of large amount of atoms to emergent properties~\cite{adcock2006molecular,hospital2015molecular,hollingsworth2018molecular,schlick2021biomolecular}.
However, despite decades of development, its application to large biomolecular systems is hampered by its computational cost. As atomistic MD computes the interactions of all biomolecular and solvent atoms and advances simulation time by a few femtoseconds per step, it is often computationally prohibitive to simulate timescales relevant for biomolecular dynamics. This has led to the creation of coarse-grained (CG) MD models which  use far fewer effective atoms and can employ effectively longer simulation timesteps, resulting in large speedups~\cite{clementiCOSB2008,noid2013perspective,liwo2021theory,jin2022bottom,noid2023perspective,borges2023pragmatic,marrink2023two}.
For example, while atomistic MD could model a solvated protein by explicitly representing each atom in the protein and surrounding water, CG MD could do so by considering only the atoms in the protein backbone~\cite{noid2013perspective}.

Unfortunately, while atomistic MD has achieved high accuracy~\cite{robustelli2018developing,best2019atomistic}, CG MD has struggled to do so~\cite{noid2023perspective,marrink2023two,borges2023pragmatic}. MD propagates each particle in time according to the forces on each atom -- the negative derivatives of a potential energy function, which is generally called a force-field. Especially for biomolecular systems, atomistic force-fields have been improved over decades using a combination of quantum mechanical calculations and experimental data. And despite limitations in their functional form, modern force-fields have reached a level of high predictiveness in terms of metastable structures and free energy differences, if sufficient sampling is used~\cite{shaw2010atomic,voelz2010molecular,lindorff2011fast,plattner2017complete,robustelli2022molecular,best2019atomistic}. %
However, CG force-fields are still far away from their atomistic counterparts in generalization and accuracy. Recent progress has suggested that machine learning (ML) may be central to achieving this goal~\cite{Wang2019,durumeric2023machine,majewski2022machine,charron2023navigating}, as ML can learn crucial many-body interactions which are often missing from traditional CG force-fields~\cite{wang2021multibody,zaporozhets2023multibody}. %

Many methods exist for representing and training CG force-fields~\cite{jin2022bottom,noid2023perspective,borges2023pragmatic}. 
Training approaches typically either pursue a top-down approach, in which one matches the values of macroscopic quantities~\cite{marrink2023two,borges2023pragmatic}, or a bottom-up approach that aims to reproduce detailed behavior present in a corresponding atomistic model~\cite{jin2022bottom,noid2023perspective}. Here we focus on the bottom-up approach following multiscale coarse-graining (also known as ``variational force-matching'')~\cite{izvekov2005multiscale,izvekov2005liquidmultiscale,Noid2008,lu2010efficient,lu2012multiscale}.
In terms of the CG force-field representation, it is nontrivial how to design the functional form of the force-field itself to be sufficiently expressive yet practical for a given application.
Recent attempts have borrowed function representations from ML~\cite{durumeric2023machine} (\textit{e.g.}, neural networks~\cite{bishop2024deep}), 
an approach that has seen preliminary success in modelling biomolecules~\cite{lemke2017neural,Wang2019,Husic2020,wang2021multibody,Ding2022coarsegrained,majewski2022machine,flowmatching2023,chennakesavalu2023ensuring,kraemer2023statistically,wellawatte2023neural,airas2023transferable,arts2023two,charron2023navigating}. 
However, learning the corresponding force-field has required either substantial computation or a significant amount of data~\cite{durumeric2023machine}.
The resources necessary for the creation of machine-learned coarse grained (MLCG) 
force-fields represent a significant barrier to the development of the next generation of CG models which could rival the accuracy of atomistic MD. 

Here we present a strategy for training bottom-up neural-network CG force-fields using reference atomistic MD simulation data that is computationally efficient and reduces data requirements by a factor of 100. This is made possible by unifying denoising score matching~\cite{hyvarinen2005estimation,vincent2011connection,DenoisingScoreMatchig} (the key technology underlying denoising diffusion models~\cite{sohl2015deep,ho2020denoising,song2020score}, a popular type of ML generative model) with the learning of MLCG force-field parameterization based on the ``force-matching" approach~\cite{Noid2008}.
This connection in turn suggests novel modifications for both classes of models. The corresponding methodologies are implemented in public code-bases and demonstrated by creating MLCG force-fields of the Trp-Cage~\cite{barua2008trp} and NTL9~\cite{cho2004ntl9} proteins using minimal amounts of training data.

\section{Combined force matching and denoising for learning coarse-grained force-fields}

\paragraph{\label{sec:cg}Learning from Forces}

We train a CG neural network potential $\cgff_\cgffparam(\cgr)$ with neural network parameters $\theta$ as a function of the CG coordinates, $\cgr$. 
Force-matching minimizes the mean squared difference between the atomistic forces $\fgforce(\fgr)$ at configuration $\fgr$, projected onto the CG space, and the forces of the CG force-field. 
The CG coordinates $\cgr$ are usually defined as a linear mapping of the atomistic coordinates:~$\cgr = \cmap\fgr$. 
The effective CG force-field which produces the same equilibrium distribution as the corresponding atomistic system, referred to as the ``many body potential of mean force'' ($\mbpmf$), minimizes the loss function~\cite{Noid2008,noid2013perspective}:
\begin{equation}
        \label{eq:fm}
        \forceresidual (\cgffparam) 
         = 
        \raverage{
        \Bigl\|
             \cgforce_\cgffparam(\cgr)
             - \fmap \fgforce (\fgr)
        \Bigr\|_2^2}_{\fgr} 
\end{equation}
where 
$\cgforce_\cgffparam(\cgr) = -\grad \cgff_\cgffparam(\cgr)$ is the CG force-field generated by the learned CG potential,
$\langle \cdot \rangle_{\fgr}$ denotes an ensemble average over the equilibrium distribution of the atomistic system,
and $\fmap$ transforms atomistic forces into forces on the CG particles~\cite{Ciccotti2005,Noid2008,kraemer2023statistically}.

In order to accurately approximate the potential of mean force $\mbpmf$ implied by the atomistic simulation, a flexible functional form is needed for $\cgff_\cgffparam(\cgr)$, and neural networks have shown to be a formidable choice~\cite{durumeric2023machine}.
However, training neural network CG force-fields for biomolecular systems requires a lot of data, e.g., many diverse configurations with force labels~\cite{kraemer2023statistically}.
Here we propose a much more data-efficient approach that leverages not only the force labels but also the distribution of configurations $\cgr$ in the training data while avoiding repeated long simulation of the CG model~\cite{jin2022bottom,noid2023perspective,Ding2022coarsegrained,airas2023transferable,thaler2021learning,thaler2022deep}.

\paragraph{Learning from Noise}
The proposed training modifications extend denoising score matching~\cite{DenoisingScoreMatchig}, which is the central idea in Denoising Diffusion Models (DDMs) 
\cite{sohl2015deep,ho2020denoising,song2020score},
a class of machine-learning algorithm trained to transform noise variates
to approximate the distribution of training data. Analogous to distribution-based force-field learning~\cite{jin2022bottom,noid2023perspective}, DDMs learn the distribution present in a data sample without pre-recorded force information; instead, this information is learned through the addition and removal of noise.

During DDM training, data is first corrupted to different extents with additive noise, with maximal corruption giving an uninformative prior distribution and no corruption corresponding to the original data distribution.
This corruption process is captured by introducing a conditional density $\ddmdensity(\ddmR|\ddmr)$, which describes noising a training data sample $\ddmr$ to obtain the noised data $\ddmR$. We can interpret the $-\log\ddmdensity(\ddmR|\ddmr)$ as a ``noise energy" and its negative gradients as ``noise forces'', $\cgforce_{\ddmdensity} = \grad_{\ddmR} \log \ddmdensity(\ddmR|\ddmr)$. A neural network $\scorecandidate_{\cgffparam}(\ddmR)$, called ``score network", %
is trained to predict the noise forces by minimization of a loss function.
This loss function is defined as a sum of individual denoising score matching losses  for each distortion level. The loss for a single distortion level can be interpreted as force matching (Eq.~\ref{eq:fm}) further averaged over the noise variables:
\begin{equation}
\label{eq:ddmtrain}
    \forceresidual_{\text{DSM}}(\cgffparam) =
    \left\langle
    \expect_{\ddmdensity}
    \Bigl\|
    \scorecandidate_{\cgffparam}(\ddmR) -
       \cgforce_{\ddmdensity}(\ddmR,\ddmr)
    \Bigr\|^2_2 %
    \right\rangle_{\ddmr},
\end{equation} 
where the brackets $\langle \cdot \rangle_{\ddmr}$ denote expectation over the data distribution via $\ddmr$ and $\expect_{\ddmdensity}$ is the expectation over the noise variables  $\ddmR$.
DDMs have been shown to be successful at various tasks~\cite{yang2023diffusion},
including modeling CG protein conformational landscapes by fitting the training data distribution~\cite{arts2023two,Zheng2024}.

\paragraph{Combining Noise and Forces in Coarse-Graining}
We propose to augment Eq.~\eqref{eq:fm} by adapting the denoising described in Eq.~\eqref{eq:ddmtrain} for a single noise level through the definition of a modified loss function:
\begin{equation}
    \begin{aligned}
        \label{eq:kernelfm}
        \forceresidual^\kernel (\cgffparam; \fmap) 
        =
        \raverage{
        \expect_{\kernel}
        \Bigl\|
            \cgforce_{\cgffparam}(\cgr)
            - 
            \fmap \fgforce(\cgr,\fgr)
        \Bigr\|_2^2
        }_{\fgr}
    \end{aligned}
\end{equation}
where the force $\fgforce(\cgr,\fgr)$ is now defined as sum of the atomistic MD forces and the noise forces, where energies are specified in thermal units ($\kbt = 1$):
\begin{equation}
    \label{eq:fluctforce}
    \fgforce(\cgr,\fgr)= %
        \underbrace{ -\grad \fgff(\fgr)}_{\mathrm{MD\;Forces}}
         + 
        \underbrace{\;\grad \log \kernel  (\cgr,\fgr)}_{\mathrm{Noise\;Forces}}.
\end{equation}
The noise  $\kernel(\cgr,\fgr)$ is here utilized as a probabilistic replacement to the CG map $\cmap$: as where $\cmap$ deterministically defines the CG positions $\cgr$ given the fine-grained positions $\fgr$, $\kernel$ instead describes the probability density of obtaining a given $\cgr$ conditioned on $\fgr$.

In Eq.~\eqref{eq:fm}, the force map $\fmap$ projects the atomistic forces into the  CG space. As shown previously~\cite{kraemer2023statistically}, different choices of $\fmap$s can be made for a given choice of CG coordinates. In the context of Eq.~\eqref{eq:kernelfm}, $\fmap$ also specifies how to combine distributional and atomistic force information.
$\fmap$ may be defined to include only contributions from $\kernel$, facilitating force-field optimization when no atomistic forces are available; the corresponding $\fmap$ is referred to as $\fmap_{\mathrm{noise}}$.
In the case of $\fmap_{\mathrm{noise}}$,  training in the proposed framework is equivalent to the training of a DDM:  Eq.~\ref{eq:kernelfm} reduces to Eq.~\ref{eq:ddmtrain} at a fixed noise level. On the other hand, Eq.~\ref{eq:kernelfm} reduces to Eq.~\ref{eq:fm} if a $\kernel$ concentrated at $\cmap\fgr$ is considered (see Appendix).

Critically, however, $\fmap$ may instead be defined to include contributions from both saved atomistic forces and noising, allowing the seamless combination of distributional and force-based information during training. Preliminary work has shown that combining reference force information with noising may improve performance in the low-noise stage of DDMs in two dimensions~\cite{de2024target}; the current work demonstrates that the theory of bottom-up CG can unify this information to create state of the art MLCG force-fields on complex systems with drastically reduced amounts of training data while retaining a rigorous connection to the atomistic system.

\paragraph{Improved MLCG Force-Fields}
MLCG force-fields were created by optimizing neural networks to minimize Eqs.~\eqref{eq:fm} or~\eqref{eq:kernelfm} for Trp-Cage~\cite{barua2008trp} and NTL9 (K12M mutant)~\cite{cho2004ntl9}, two proteins that have previously required millions of frames for accurate MLCG force-field parameterization~\cite{majewski2022machine,krishna2024generalized}.
Force-fields were created at a CG resolution of one site per $\alpha$-carbon and trained using batched gradient-based optimization.
When minimizing Eq.~\eqref{eq:fm}, batches were used without modification to estimate the gradient. 
However, in the case of Eq.~\eqref{eq:kernelfm}, batches were drawn and augmented with standard Gaussian noise; this corresponds to $\kernel(\cgr,\fgr)$ defined to be a Gaussian centered at $\cmap\fgr$ with diagonal covariance.
This procedure is described in Figure~\ref{fig:diagram}. When minimizing Eq.~\eqref{eq:fm}, $\fmap$ is then defined to minimize $\langle \| \fmap \fgforce \|^2 \rangle$ as in previous work~\cite{kraemer2023statistically}. When minimizing Eq.~\eqref{eq:kernelfm}, two $\fmap$ were used: 
the first combined information from both atomistic and noise-derived forces from the minimization of $\langle \expect_{\kernel} \| \fmap \fgforce \|^2  \rangle$, and the second used only noise-related information, $\fmap = \fmap_{\mathrm{noise}}$.
Noise-based data transformations were implemented in a publicly available code base (\url{github.com/noegroup/aggforce}). 

{
\begin{figure*}
\centering
\includegraphics[scale=0.85]{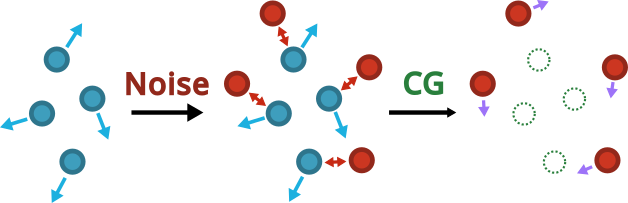}

\caption{
An illustration of the proposed training strategy. Training data consisting of ``real'' particles with associated forces (blue) are first combined with noise ($\kernel$) to add additional sites (red); these new particles interact with the original particles, changing forces throughout the system. The ``real'' particles are then systematically coarse-grained out (green) and associated with a linear combination ($\fmap$) of ``real'' and noise-derived force information (purple), providing data for subsequent force-field training.
}
\label{fig:diagram}
\end{figure*}
}

\section{Results}
We demonstrate the performance of MLCG force-fields trained with a combination of force matching and denoising on two fast-folding proteins.

\paragraph*{Trp-Cage}

{

\begin{figure*}
\centering
\includegraphics[width=5.85in]{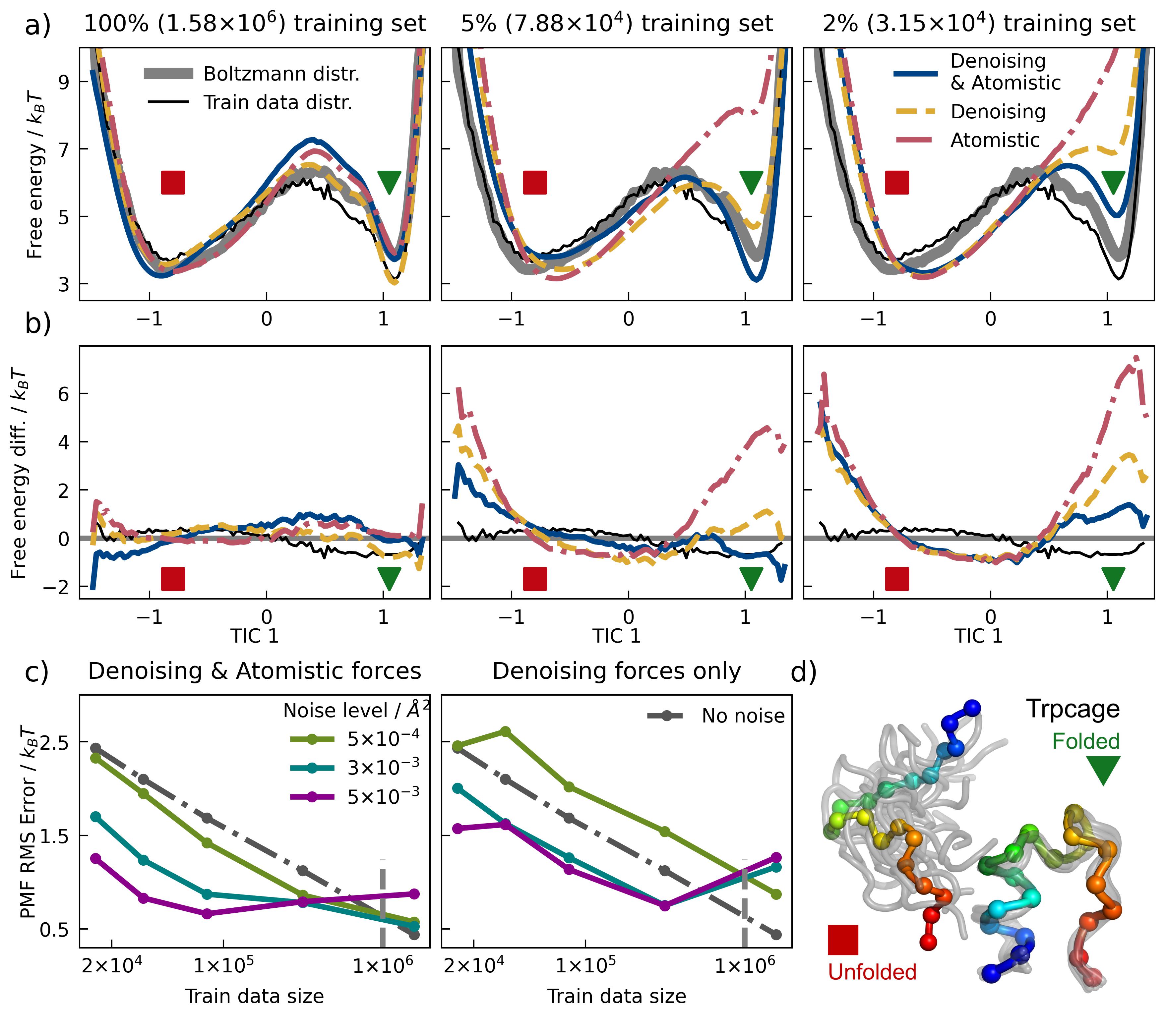}

\caption{
MLCG models of Trp-Cage.
a.~1-D free energy over the first TIC for models trained (I)~with a combination of denoising forces and mapped atomistic forces (dark blue), (II)~with denoising forces (yellow), or (III)~with atomistic forces (red) on different strides of the training dataset. The MSM-reweighted reference surface is shown in solid gray and training data distribution in thin black. Only a single noise level (0.003 $\text{\AA}^2$) is visualized. 
b.~Difference of 1-D free energy over the first TIC for models and training dataset when compared with the MSM-reweighted reference. Color scheme is the same as subplot a.
c.~Dependence of model accuracy on training size at different noise levels. 
d.~CG representation of Trp-Cage.
}
\label{fig:trpcage}
\end{figure*}
}

MLCG models of Trp-Cage~\cite{barua2008trp}, a 20 residue miniprotein, were optimized using training sets of varying sizes and $\kernel$s of different variances to investigate the data efficiency of the proposed learning procedures. 
Training samples were extracted from adaptively sampled short atomistic MD simulations~\cite{majewski2022machine} at 350$K$ that do not distribute according to the many body potential of mean force (PMF); for model validation, this data was reweighted using a Markov State Model (MSM)~\cite{prinz2011markov}.
Model accuracy was quantified using low dimensional free energy surfaces (FESs) generated by simulating the CG force-field using unbiased molecular dynamics. FESs were calculated along a low-dimensional projections capturing the folding-unfolding process as revealed by TICA~\cite{Perez_JChemPhys2013} on the atomistic trajectories (Fig.~\ref{fig:trpcage}).

Models trained on atomistic forces without noise using Eq.~\eqref{eq:fm} recover the reweighted FES (despite the biased training set) with sufficient training data (1.6M samples, Fig.~\ref{fig:trpcage}a left panel). 
In contrast, minimizing Eq.~\eqref{eq:kernelfm} using only noise information produces samples biased towards the distribution of training data (Fig.~\ref{fig:trpcage}a left panel). In this high-data regime, combining these sources of information by minimizing Eq.~\eqref{eq:kernelfm} with an optimized $\fmap$ results in similar accuracy as traditional force-matching;
however, this approach greatly increases performance at lower training set sizes, yielding significantly more accuracy than traditional force-matching when less than 1M samples are available.
Strikingly, training with only noise information is 
as accurate (for noise variance $5\times10^{-4}$ \AA$^2$), or even 
more accurate (for noise variance 0.003 and 0.005 \AA$^2$), than conventional force-matching (Eq.~\eqref{eq:fm}) in this low data regime (Fig.~\ref{fig:trpcage}a middle and Fig.~\ref{fig:trpcage}c right panel); moreover, this data efficiency is further increased by including force information.
We note that high noise levels ($>0.01$ \AA$^2$) can be detrimental, 
perhaps due to breakdown of the prior energy terms in MLCG training (see Appendix for details).

\begin{figure*}
\centering
\includegraphics[width=5.853in]{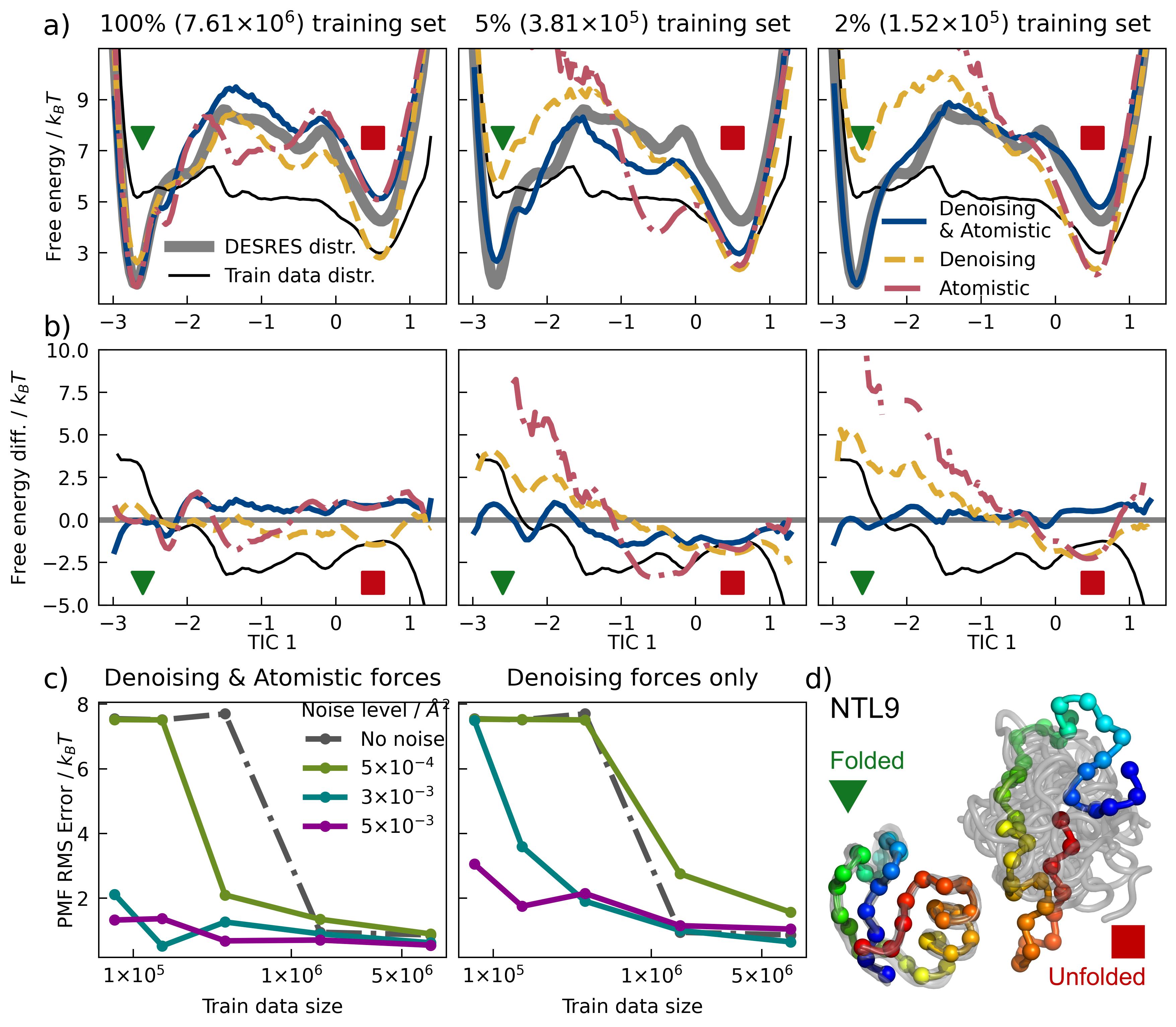}
\includegraphics[width=5.44in]{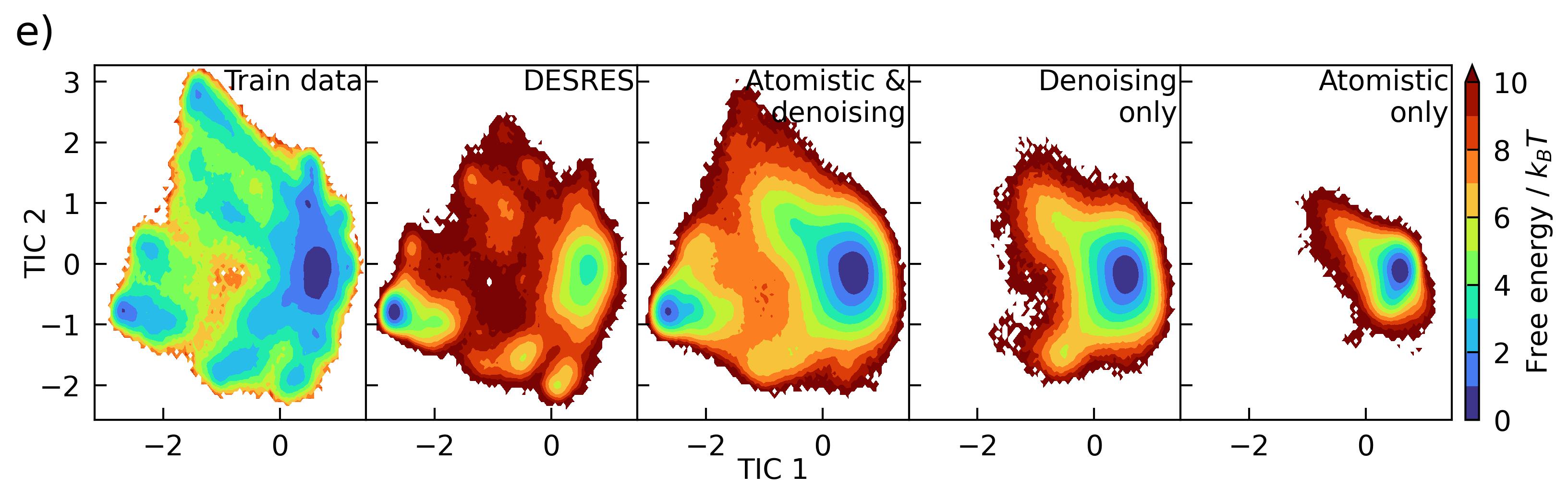}
\caption{MLCG model of NTL9.
a.~1-D FES over the first TIC for models trained (I)~with a combination of denoising forces and atomistic forces (dark blue), (II)~with denoising forces (yellow) or (III)~with atomistic forces (red) on different strides of the training dataset. The reference FES is shown in solid gray and training data distribution in thin black. Only noise level 0.003 $\text{\AA}^2$ is shown. 
b.~Difference of 1-D free energy over the first TIC for models and training dataset when compared with the reference. Color scheme is the same as subplot a.
c.~Dependence of model accuracy on training size and noise levels. 
d.~CG representation of NTL9 in its folded and unfolded states.
e.~2-D FESs for NTL9 reference simulations and MLCG models trained with 1\% of the training data. Noise level is 0.003 \AA$^2$ when noise information is used.
}
\label{fig:ntl9}
\end{figure*}

\paragraph*{NTL9}
The proposed training methodologies were additionally validated on 
NTL9 (K12M)~\cite{cho2004ntl9}, a 39 residue protein. As in the case of Trp-Cage, models were optimized using training sets of varying sizes and $\kernel$s of different variances. 
NTL9 represents a significantly more challenging learning target than Trp-Cage, featuring a slower folding-unfolding transition and several folding intermediates~\cite{voelz2010molecular, lindorff2011fast,Schwantes_JChemTheoryComput2013} (Fig.~\ref{fig:ntl9}d). 
Training samples were similarly extracted from adaptively sampled short atomistic MD simulations~\cite{majewski2022machine} at 350$K$ that do not reflect the many body PMF; in this case, creation of a converged MSM was not feasible (see Methods). Instead, millisecond-length trajectories generated at 355$K$ by D.E. Shaw Research (DESRES) using the Anton computer with the same force-field and molecular topology were utilized as an equilibrium reference~\cite{lindorff2011fast}.
The distribution of samples differs significantly between the training data and the reference Anton simulations (Fig.~\ref{fig:ntl9}a, \ref{fig:ntl9}e). We note that, as revealed by Ref.~\cite{Schwantes_JChemTheoryComput2013,Schwantes_BiophysJ2016}, even these long Anton simulations may deviate from the underlying many body PMF for minor states.

Similar to Trp-Cage, models trained using traditional force matching and a combination of force and noise information accurately approximate the distribution of the reference data despite the biased training data; in comparison, those trained using only noise information are biased towards the training distribution (Fig.~\ref{fig:ntl9}a). We note that models trained using only noise information still exhibit a bias towards the correct equilibrium distribution, while in principle should reproduce the distribution of the training data. This bias may be perhaps induced by the MLCG force-field architecture. At lower training sizes, the combination of force and noise information greatly increases accuracy.
Strikingly, at intermediate noise levels, the combined model can produce a usable MLCG force-field with only 1\% of the whole training set (76k frames, bottom left of Fig.~\ref{fig:ntl9}c).
We note that the PMF error saturates at 7.5$\kbt$ for the other models (noise only or forces only) at lowest training set sizes, as this corresponds to a model completely destabilizing the folded state and folding intermediates (Fig.~\ref{fig:ntl9}e).

\section{Discussion}
MLCG MD represents a promising alternative to atomistic simulation~\cite{jin2022bottom,charron2023navigating}. 
However, training represents a significant barrier to applying MLCG force-fields to biomolecules~\cite{durumeric2023machine}: Existing approaches have required large amounts of training data~\cite{kraemer2023statistically,charron2023navigating}, repeated converged simulations of the MLCG model being trained~\cite{thaler2021learning,thaler2022deep}, or modified sampling of the atomistic system under study~\cite{Ding2022coarsegrained,airas2023transferable}. 
These difficulties have significantly hampered the development of MLCG models relative to their atomistic counterparts~\cite{unke2021machine,durumeric2023machine}.

The present work represents a significant step forward by substantially reducing the amount of data required to efficiently train biomolecular MLCG models on atomistic MD trajectories. 
When force information has been recorded during atomistic MD, the proposed method significantly improves upon the data efficiency of traditional force matching while retaining its ability to correctly infer the underlying Boltzmann distribution from unconverged simulations. 
In the absence of force information, the proposed approach targets the sampled distribution with higher data efficiency than traditional force matching. 
In both cases no simulation of the MLCG force-field is required during optimization, representing a significant improvement over previous distribution-based training approaches~\cite{thaler2021learning,thaler2022deep}. The proposed methodological advances have been developed in a publicly available code base.

The proposed training approach is based on a well-defined reinterpretation of the theory underpinning systematic CG~\cite{Noid2008} and DDMs. This mathematical foundation allows the transparent application of existing thermodynamic frameworks to the proposed models
~\cite{jin2022bottom,noid2023perspective} and provides a straightforward link between CG force-field parameterization and DDMs. These connections may provide avenues for improving the efficiency of DDMs focused on conformational distributions~\cite{arts2023two}, and may more generally serve as a framework utilizing probabilistic maps with CG force-fields. Since the relationship between the derived forces and distribution of configurations remains valid, the proposed improvements may be directly combined with other approaches for MLCG force-field parameterization~\cite{lemke2017neural,Ding2022coarsegrained,majewski2022machine,flowmatching2023,chennakesavalu2023ensuring,airas2023transferable,duschatko2024thermodynamically}. 

Many questions about MLCG force-fields remain open. While unconverged, the training data used in this study captured a substantial level of conformational diversity; as a result it remains an open question how the proposed methods may perform on data with limited diversity, e.g., whether a model can recover states absent from data. Furthermore, while the proteins considered this work are challenging targets for MLCG force-fields~\cite{arts2023two,majewski2022machine}, they do not represent a biologically relevant application; the effect of the proposed procedures on complex CG force-fields~\cite{charron2023navigating} remains to be investigated.
Nevertheless, the proposed approach represents a striking gain in data efficiency over previous work~\cite{majewski2022machine}, greatly increasing the applicability of a promising class of physically informed ML models. 

\section{Methods}

\subsubsection{Optimizing CG Force-Fields}
MLCG force-fields were optimized to minimize their force prediction error on CG conformations. 
The CG resolution was defined by the removal of all atoms except for the alpha carbons of each amino acid; this is algebraically encoded in $\cmap$ or $\kernel$, depending on the training equation. CGSchNet
(a modified SchNet~\cite{schutt2018schnet} graph neural network plus prior energy terms) 
with two interaction blocks was used to represent all CG force-fields~\cite{Husic2020}. Auto-differentiation was used to obtain forces on each CG bead.

When optimizing force-fields using traditional force matching,  established procedures were used~\cite{Wang2019,Husic2020,kraemer2023statistically}. Positions and forces were drawn from atomistic MD simulations and mapped to the CG resolution using preselected $\cmap$ and $\fmap$ matrices and used to generate batched gradient updates using the ADAM optimizer.
When minimizing Eq. \eqref{eq:kernelfm}, the gradient update was modified as follows. 
Batches were first drawn and mapped to the non-noised alpha carbon resolution. 
Second, 0-mean Gaussian noise with diagonal covariance was added to the positions; collectively, this corresponds to $\kernel(\cgr,\fgr)$ defined to be a Gaussian centered at $\cmap\fgr$. 
Third, forces were modified according to Eq.~\eqref{eq:fluctforce} with a preselected $\fmap$.
$\fmap$ was defined either to minimize $\langle \expect_{\kernel} \| \fmap \fgforce \|^2  \rangle$ or, in the case of noise-only forces, directly defined via Eq.~\ref{eq:noisefmap} (see Appendix).

To investigate the data efficiency of different training strategies and noise levels, training and validation set size was varied. Smaller data sets were created from the full data set via striding, ensuring that the conformational distribution did not change. For each training condition, a representative model was selected for MD analysis using the following criteria. After the initial decrease of the validation loss, if a consistent uptick of the validation loss over multiple epochs was observed, the model at the lowest reached validation loss was selected. In case of no obvious uptick (e.g., for large training sizes with denoising information), training was stopped at a certain number of epochs (e.g., 200 for NTL9 models on full training data) and used for simulation analyses. Note that there were cases where the model quality seemed to be sensitive to stochastic fluctuations during training
or deteriorated over epochs despite validation losses continuing to gradually decrease, 
especially for NTL9 models at intermediate noise levels (e.g., for a variance of 0.003 \AA$^2$). If simulation instability was observed, an earlier epoch was selected. More details can be found in the Appendix.

\subsubsection{Optimizing $\fmap$}

All force maps were optimized using code implemented in a publicly available repository at \url{github.com/noegroup/aggforce}; force maps including noise were generated via the \verb|stagedjoptgauss_map| and \verb|stagedjslicegauss_map| methods. Supporting libraries include numpy~\cite{harris2020array}, jax~\cite{jax2018github}, pandas~\cite{reback2020pandas}, scipy\cite{2020SciPy-NMeth}, and OSQP~\cite{osqp,osqp-infeasibility}.
Force maps were optimized using an L2 regularization of $10^{3}$ for atomistic contributions and $5$ for post-map noise contributions (see Appendix); forces were considered in units of $\text{kcal}/(\text{mol}\cdot\text{\AA})$. Force map optimization was performed using 350000 and 500000 MD frames for Trp-Cage and NTL9, respectively.

\subsubsection{Reference Atomistic MD Simulations}
MLCG models were trained and validated against solvated all-atom MD simulations. For training, a dataset generated on GPUGRID~\cite{buch2010high} was used~\cite{majewski2022machine}, which utilized thousands of short simulations seeded with an adaptive sampling strategy.
Collectively, this dataset provides approximately 2M frames of coordinates and forces from 3940 simulations (50$ns$ each) for Trp-Cage and 9.5M frames from 47599 trajectories (20$ns$ each) for NTL9.
The simulation procedure and data availability are described in Ref.~\cite{majewski2022machine}.
Owing to the long timescale of major conformational transitions, the data distribution deviates from the true Boltzmann distribution for the all-atom systems. For Trp-Cage, we obtain the reference FES with MSM reweighting as in Ref.~\cite{kraemer2023statistically}. For NTL9, due to difficulties in obtaining implied timescales consistent with Markovian dynamics~\cite{prinz2011markov}, data generated using the Anton supercomputer by D.E. Shaw Research was used as reference for the equilibrium distribution. This data comprises four long simulations (2.9$ms$ in aggregate time) with the same force-field and solvation setup~\cite{lindorff2011fast}. The  ANTON simulations were performed at a temperature of 355K instead of 350K of the training data from GPUGRID, implying that the folded state is expected to be slightly destabilized relatively to the targeted many body potential of mean force.

\subsubsection{CG MD Simulations}
Trained CG models were evaluated by performing 100 MD simulations at the same temperature as the training data (350K). Simulation used the Langevin thermostat with a friction coefficient of 1 ps$^{-1}$ and time step of 2 fs. For Trp-Cage each trajectory was simulated for 5M time steps, while for NTL9 at least 10M steps. The simulation time was sufficiently long to allow several observations of state transitions along the first TIC, and thus close to convergence for the slowest dynamics. Note that a small number of NTL9 models exhibited integration instability; this was addressed by re-initialization of the corresponding simulation. 
This might be attributed to the imperfect functional form and fitting of prior terms based on Gaussian-noised coordinates statistics. For implementation details the readers are referred to Ref.~\cite{kraemer2023statistically}.

\section{Acknowledgements}
The authors thank Andreas Kr\"{a}mer and Atharva Kelkar for helpful discussions and Jacopo Venturin for Markov State Model information.
We gratefully acknowledge funding from the Deutsche Forschungsgemeinschaft
DFG (SFB/TRR 186, Project A12; SFB 1114, Projects B03, B08, and A04; SFB 1078, Project C7), the National Science Foundation (PHY-2019745), and the Einstein Foundation Berlin (Project 0420815101), and the computing time provided on the supercomputer Lise at NHR@ZIB as part of the NHR infrastructure.

\clearpage

\appendix

\setcounter{table}{0}
\renewcommand{\thetable}{A\arabic{table}}
\setcounter{figure}{0}
\renewcommand{\thefigure}{A\arabic{figure}}

\section{}
We first provide an extended theoretical description of the noised many body potential of mean force (mb-PMF) and then discuss additional numerical results on proteins Trp-Cage and NTL9.

\subsection{Theory}

The equations in the main text and this appendix apply to a system of $\fgn$ atoms with coordinates $\fgr\in V \subset \reals^{3\fgn}$ in the canonical ensemble.
The CG resolution is defined by the full rank matrix $\cmap$ which projects the atomistic coordinates $\fgr$ to their coarse-grained (CG) counterparts $\cmap\fgr = \cgr \in \reals^{3\cgn}$ comprising $\cgn$ ``beads'' with $\cgn \ll \fgn$. Both the atomistic and CG force-fields are described here in thermal units ($\thermb = 1$), and the two systems are set to have the same temperature in practice. We only consider configurational consistency (and not momentum consistency)~\cite{Noid2008}.
 
\subsubsection{Noise and the mb-PMF}
As stated in the main text, combining noise with atomistic force information can be described using a function referred to as the noised mb-PMF.
This surface may be mathematically understood using tools central to bottom-up coarse graining, which typically creates CG force-fields approximating the (unnoised) mb-PMF~\cite{jin2022bottom,noid2023perspective}. The unnoised mb-PMF ($\mbpmf$) is defined (up to a constant) as:  
\begin{equation}\label{eq:mbpmf}
     \mbpmf(\cgr) = - \log \int \delta [\cgr - \cmap\fgr] 
     \exp \left[ - \fgff(\fgr) \right] \md \fgr
\end{equation}
where integration is applied over the entire domain of the corresponding probability distribution. 
Eq.~\eqref{eq:mbpmf} may be modified by replacing the Dirac $\delta$ with a kernel function $\kernel$. Unlike $\delta$, $\kernel$ is not defined solely as a distribution, but is instead a positive scalar-valued function with well-defined log gradients with respect to all arguments.
Given its relationship to $\delta$, intuitive $\kernel$ are concentrated where $\cgr$ is close to $\cmap\fgr$ for a predefined $\cmap$; in the experiments of this work, $\kernel$ is defined to be a multivariate Gaussian density over $\cgr$ centered at $\cmap\fgr$ with a covariance matrix proportional to the identity matrix. 
This substitution results in a ``smoothed'' or noised variant of the mb-PMF 
 ($\mbpmf^\kernel$) defined as:
\begin{equation}\label{eq:softmbpmf}
    \mbpmf^\kernel(\cgr) = - \log \int \kernel [\cgr,\fgr] \exp \left[ -\fgff(\fgr) \right] \md \fgr.
\end{equation}
Critically, as $\kernel$ is strictly positive, the integration against $\kernel$ present in Eq.~\eqref{eq:softmbpmf} may be itself understood as a traditional coarse-graining problem, where our CG configurational map is defined to be $(\fgr,\cgr) \mapsto \cgr$ (isolation of $\cgr$):
\begin{equation}\label{eq:softhard}
    \int \kernel [\cgr,\fgr] \exp \left[ -\fgff (\fgr) \right] \md \fgr
    =
    \int \delta [\cgr - \cgr'] \exp \left[ -\fgff(\fgr) + \log \kernel (\cgr',\fgr)\right] \md \fgr \md \cgr'.
\end{equation}
These equations interpret the effect of $\kernel$ through an induced force-field term $\log \kernel (\cgr',\fgr)$ with $\cgr'$ part of an extended configurational state. This view may in turn be combined with established literature to understand rigorous requirements on $\kernel$~\cite{ciccotti2008projection, kalligiannaki2015geometry}; we note that for all $\kernel$ discussed in the main text, the corresponding $\log \kernel$ are harmonic potentials in the CG space, centered on $\cmap\fgr$. As Eq. \eqref{eq:softmbpmf} illustrates, traditional non-noised coarse-graining expressions are regained via the limiting case of Gaussian $\kernel$ with vanishing variance, although Eq. \eqref{eq:softhard} then becomes undefined.

As Eq.~\eqref{eq:softhard} represents a typical coarse-graining operation, it may be analyzed through the lens of force-matching (multiscale coarse-graining)~\cite{Noid2008}, giving rise to Eq.~\eqref{eq:kernelfm}. However, it is important to remark that the gradient operation typically applied to $\fgff$ to define forces must instead be applied to $\fgff-\log \kernel$ and taken with respect to both $\fgr$ and $\cgr$; this operation defines forces on the $\kernel$-induced particles $\cgr$ and modifies the forces on the ``real'' particles $\fgr$; in the main text, the corresponding gradient domain is made clear by context. The combination of the implied CG map and modified forces, when combined with existing framework relating transformed atomistic forces to their CG counterparts~\cite{Ciccotti2005,Noid2008,ciccotti2008projection,kraemer2023statistically}, implies the various $\fmap$s utilized in the main text. In particular, mapping the forces to retain only the forces on the configurationally preserved particles results in $\fmap_{\text{noise}}$, as $\fgff$ is \emph{not} a function of the preserved particles.
\begin{multline}\label{eq:noisefmap}
\fmap_{\mathrm{noise}}  \fgforce(\cgr,\fgr) =
    \fmap_{\mathrm{noise}}  
    \left[ -\grad \fgff(\fgr) + \grad \log \kernel (\cgr,\fgr) \right] \\
    = -\grad_{\cgr} [\fgff(\fgr) - \log \kernel (\cgr,\fgr)]
    = \grad \log \kernel (\cgr,\fgr)
\end{multline}
The extended phase space viewpoint described in Eq.~\eqref{eq:softhard} allows corrections related to constrained bonds to be incorporated in a straightforward way using existing work~\cite{Ciccotti2005,kraemer2023statistically}; all derived $\fmap$ used in this work respect these conditions.
We further note that analysis related to the thermodynamic state may be performed as long as choices are made with regard to the state-dependence of $\kernel$~\cite{jin2022bottom,noid2023perspective}. Replacement of Dirac deltas with Gaussian functions has been historically used in the derivation of methods typical to enhanced sampling such as umbrella sampling~\cite{torrie1977nonphysical} and adiabatic free energy dynamics~\cite{rosso2002use,maragliano2006temperature}, providing intriguing avenues for further analysis.

As where minimization of Eq.~\eqref{eq:fm} results in $\mbpmf$, minimization of Eq.~\eqref{eq:kernelfm} results in $\mbpmf^\kernel$. These are generally not the same function; however, as shown in the main text, $\kernel$ may be chosen such that they are very similar. In this view the expressions discussed in the main text introduce a \emph{bias} into the learning objective~\cite{hastie2009elements}; however, as numerically demonstrated, finite-sample training procedures derived from these biased objectives may better reproduce the metastable behavior in the unnoised mb-PMF than their unmodified counterparts. Furthermore, while the mb-PMF may seem the ultimate goal due to its amenability to theoretical analysis~\cite{jin2022bottom,noid2023perspective}, we note that the noised mb-PMF may also represent an important object in its own right if it corresponds to a smoother surface allowing for accelerated sampling. This topic will be explored in future work.

If $\kernel(\cgr,\fgr):=\phi[\cgr-\cmap\fgr]$ for some function $\phi$ obeying suitable boundary conditions, algebraic operations typically directly performed on $\delta$ in Eq.~\eqref{eq:mbpmf} (e.g., chain rule and integration by parts~\cite{Ciccotti2005,Noid2008}) may similarly be performed on $\kernel$ in Eq.~\eqref{eq:softmbpmf}. This connection may be used to create training forces corresponding to the noised ensemble that do not utilize the log gradients of $\kernel$, but rather only use contributions from $\fgff$. 
More broadly, as in the case of unnoised force-matching~\cite{kraemer2023statistically}, many possible $\fmap$ operators are compatible  with the integration in Eq.~\eqref{eq:softhard}. Certain maps, such as $\fmap_\text{noise}$, only utilize force information associated with a particular set of particles; others, such as those used when combing force and noise information, may be selected for favorable learning properties.

\subsubsection{Ensemble Averages via Noising}

In the case of traditional force-matching, the ensemble average in Eq.~\eqref{eq:fm} is approximated using samples from atomistic MD. However, the expectation in Eq.~\eqref{eq:kernelfm} over the extended system defined in
Eq.~\eqref{eq:softhard} is not directly available from our reference simulations.
In this work we approximate these extended ensemble averages by combining atomistic MD with ancestral sampling~\cite{bishop2006pattern}. 
In general, ancestral sampling draws variates from a joint probability density function $f(x,y)$ by first sampling $x$ from its marginal distribution $f_X :=\int f(\cdot,y) \md y$ and then sampling $y$ via its conditional probability $f(x,\cdot)/f_X(x)$. This may be applied to the extended ensemble by first noting that said ensemble is governed by probability density \mbox{$\density \propto \kernel \exp [ -\fgff ] $}. 
If $\int \kernel (\cgr,\fgr) \md \cgr$ is independent of $\fgr$, the marginal density over $\fgr$ remains proportional to $\exp[-\fgff]$ and we may interpret $\kernel(\cgr,\fgr)$ to be proportional to the conditional probability of $\cgr$ given $\fgr$. In this case we may use the following procedure to generate samples from the extended ensemble:
\begin{enumerate}
    \item Generate $n$ samples from $\exp[-\fgff]$, denoted $\fgr_i$
    \item Given each sample $\fgr_i$, sample $\cgr_i \sim \kernel(\cdot,\fgr_i)$ to create $(\fgr_i,\cgr_i)$
\end{enumerate}
When $\exp[-\fgff]$ corresponds to the equilibrium density of the atomistic molecular dynamics simulation and $\kernel(\cgr,\fgr) \propto \exp[- (\cmap\fgr-\cgr)^{\intercal} \covmat^{-1} (\cmap\fgr-\cgr) ]$ for suitable covariance matrix $\covmat$, this procedure describes the process of injecting Gaussian noise: We first generate atomistic configurations
and subsequently combine them with Gaussian noise to create an extended ensemble containing both the noised and unnoised variables.

Correlation-based force-field training strategies (e.g.,~\cite{noid2007multiscale,shell2008relative}) directly utilize the distribution of configurations to determine a force-field. In the case of unnoised coarse-graining, mapping samples from an equilibrium atomistic simulation produces configurations drawn from the mb-PMF. When applied to equilibrium atomistic samples, the above $\kernel$-based sampling procedure similarly produces samples which, when mapped to isolate the $\kernel$-induced particles, correspond to the noised mb-PMF. As a result, correlation-based training techniques may be combined with noised samples in the same way as is done for non-noised coarse-graining~\cite{flowmatching2023}. This property allows the prior energy terms present in the utilized force-field architecture~\cite{Husic2020} to be calculated via histograms of noised samples at each noise level. As these prior terms are represented by fixed functional forms, high noise may create distributions that are difficult for the priors to approximate; this may be the cause of breakdown at higher noise levels. The strong relationship between noise-derived forces and the resulting empirical correlations may provide interesting extensions to previous unifications of force and correlation information~\cite{noid2007multiscale,mullinax2009generalized,rudzinski2011coarse}. Our numerical results on mixing force and noise information demonstrate that when applied to weakly non-equilibrated samples from atomistic simulation, low variance noise combined with atomistic forces information approximately maintains the ability of force-matching to learn the underlying mb-PMF from non-equilibrium samples. Partial insight into this phenomenon can be seen by log-differentiating Eq.~\eqref{eq:softmbpmf} in the same manner as in previous work~\cite{Ciccotti2005,Noid2008}, showing that locality in the definition of $\kernel$ in turn induces locality in contribution of atomistic forces.

\subsection{Additional Numerical Results}
\subsubsection{Model training}
MLCG models were calibrated on coordinates and forces generated via all-atom simulations. Training was performed either using the same procedure as in previous work~\cite{kraemer2023statistically}  (Eq.~\eqref{eq:fm}),
  against denoising scores (Eq.~\eqref{eq:noisefmap}), or using a linear combination of force and noise information (Eq.~\eqref{eq:fluctforce}).
When $\kernel$ was used, coordinates were noised and forces
were replaced by the corresponding terms during the creation of each batch and used to define the least-squares regression loss.
For an objective comparison of the impact of the different noise levels and force-definition strategies, other factors are held constant; these factors included model hyperparameters (e.g., SchNet architecture, the fitting procedure of prior terms), training setups (e.g., batch size, learning rate) as well as the training-validation splits.

\subsubsection{Model selection}

A criterion is needed to decide which training epoch  should represent the best model associated with each training setup.
Since different learning conditions result in different training dynamics and rates of overfitting, selecting the same epoch for models which vary only in the  definition of forces does not result in a fair comparison.
Nevertheless, we aim to use a unified approach for the model selection such as to reduce subjective bias and random variation.
We base epoch selection primarily on picking the model with lowest validation loss.
This is possible when validation loss curves exhibit an uptick; this occurs for pure force matching or low noise level training on low-data schemes.
However, for cases with abundant training data or a high noise level, no uptick of the validation loss is observed.
As a result, in these cases training was stopped at a certain number of epochs (e.g., 200 for NTL9 models on full training data) and the corresponding model was then used for simulation. 
Note that even after the validation loss plateaued, there were cases where the model quality seemed to fluctuate, or even deteriorate, during continued training despite decreasing validation losses. This was especially observed in NTL9 models at intermediate noise levels (e.g., 0.003 \AA$^2$). This issue was mitigated by selecting an earlier epoch:~checkpoints for models trained on 100\% and 20\% training data were chosen such that the number of optimizer updates was equal to that utilized for the epoch selected for training on 5\% data.

In Table~\ref{tab:trpcage_epoch_appendix} and~\ref{tab:ntl9_epoch_appendix} we summarize the epoch numbers selected for the analyses in this work and which strategy was used. We note that it has been observed in previous work that the validation force/score matching loss alone does not guarantee the best model for simulation~\cite{fu2022forces,duschatko2024uncertainty}.
The stability of the CG model training and effective assessment of the quality of a CG force-field without extensive simulations are open challenges and future work is needed to study them systematically.

\begin{table*}
\caption{\label{tab:trpcage_epoch_appendix}Epochal checkpoints chosen for Trp-Cage. The value for force \& noise and noise-only models are on the left and right of the slash for non-vanishing noise levels, respectively.}
\begin{ruledtabular}
\begin{tabular}{cccccc}
 Selected Epochs & \multicolumn{5}{c}{Percentage of training set used / Epochal saving stride }\\
 
 Noise level (\AA$^2$) & 100\%/1 & 20\%/1 & 5\%/2 & 2\%/5 & 1\%/10 \\ 

\hline
0 & 158$^{\text{b}}$ & 17$^{\text{b}}$ & 33$^{\text{b}}$ & 19$^{\text{b}}$ & 19$^{\text{b}}$ \\
0.0005 & 199\footnote{Picked at fixed epoch after validation loss plateaued.}/84\footnote{Picked epoch with lowest validation loss.} & 49$^{\text{b}}$/31$^{\text{b}}$ & 27$^{\text{b}}$/39$^{\text{b}}$ & 29$^{\text{b}}$/19$^{\text{b}}$ & 29$^{\text{b}}$/39$^{\text{b}}$ \\
0.003 & 199$^{\text{a}}$/146$^{\text{b}}$ & 199$^{\text{a}}$/61$^{\text{b}}$ & 89$^{\text{b}}$/53$^{\text{b}}$ & 59$^{\text{b}}$/44$^{\text{b}}$ & 39$^{\text{b}}$/49$^{\text{b}}$ \\
0.005 & 199$^{\text{a}}$/121$^{\text{b}}$ & 144$^{\text{b}}$/181$^{\text{b}}$ & 155$^{\text{b}}$/49$^{\text{b}}$ & 164$^{\text{b}}$/49$^{\text{b}}$ & 119$^{\text{b}}$/59$^{\text{b}}$ \\
\end{tabular}
\end{ruledtabular}
\end{table*}

\begin{table*}
\caption{\label{tab:ntl9_epoch_appendix}Epochal checkpoints chosen for NTL9. The value for force \& noise and noise-only models are on the left and right of the slash for non-vanishing noise levels, respectively.}
\begin{ruledtabular}
\begin{tabular}{cccccc}
 Selected Epochs & \multicolumn{5}{c}{Percentage of training set used / Epochal saving stride}\\
 
 Noise level (\AA$^2$) & 100\%/1 & 20\%/1 & 5\%/5 & 2\%/5 & 1\%/10 \\ 

\hline
0 & 199\footnote{Picked at fixed epoch after validation loss plateaued.} & 899$^{\text{a}}$ & 119\footnote{Picked epoch with lowest validation loss.} & 44$^{\text{b}}$ & 29$^{\text{b}}$ \\
0.0005 & 179$^{\text{b}}$/170$^{\text{b}}$ & 971$^{\text{b}}$/449$^{\text{b}}$ & 359$^{\text{b}}$/239$^{\text{b}}$ & 69$^{\text{b}}$/74$^{\text{b}}$ & 59$^{\text{b}}$/49$^{\text{b}}$ \\
0.003 & 54\footnote{Picked with comparable training batch count as the epoch picked at stride 20.}/196$^{\text{b}}$ & 275$^{\text{c}}$/564$^{\text{b}}$ & 1104$^{\text{b}}$/1434$^{\text{b}}$ & 854$^{\text{b}}$/854$^{\text{b}}$ & 229$^{\text{b}}$/199$^{\text{b}}$ \\
0.005 & 199$^{\text{a}}$/196$^{\text{b}}$ & 899$^{\text{a}}$/823$^{\text{b}}$ & 1454$^{\text{b}}$/1434$^{\text{b}}$ & 2429$^{\text{b}}$/854$^{\text{b}}$ & 299$^{\text{b}}$/269$^{\text{b}}$ \\
\end{tabular}
\end{ruledtabular}
\end{table*}

\subsubsection{Convergence of model simulations}
For model quality assessment, we compare the FES of the CG model with the FES of corresponding all-atom simulations. The FES of the CG model is estimated by the histogram of long simulations,
which were initialized from configurations randomly selected from the atomistic data sets.

\paragraph*{Trp-Cage}
Each Trp-Cage model was simulated for 5M time steps (2 fs each) from 100 starting structures randomly sampled from the training dataset. The first 40\% of each independent trajectory (2M steps) was discarded, while the rest was aggregated and converted to a histogram; $-\log$ of the histogram frequencies corresponds to the FES as visualized in the main text.

\begin{figure*}
\centering
\includegraphics[width=6.0in]{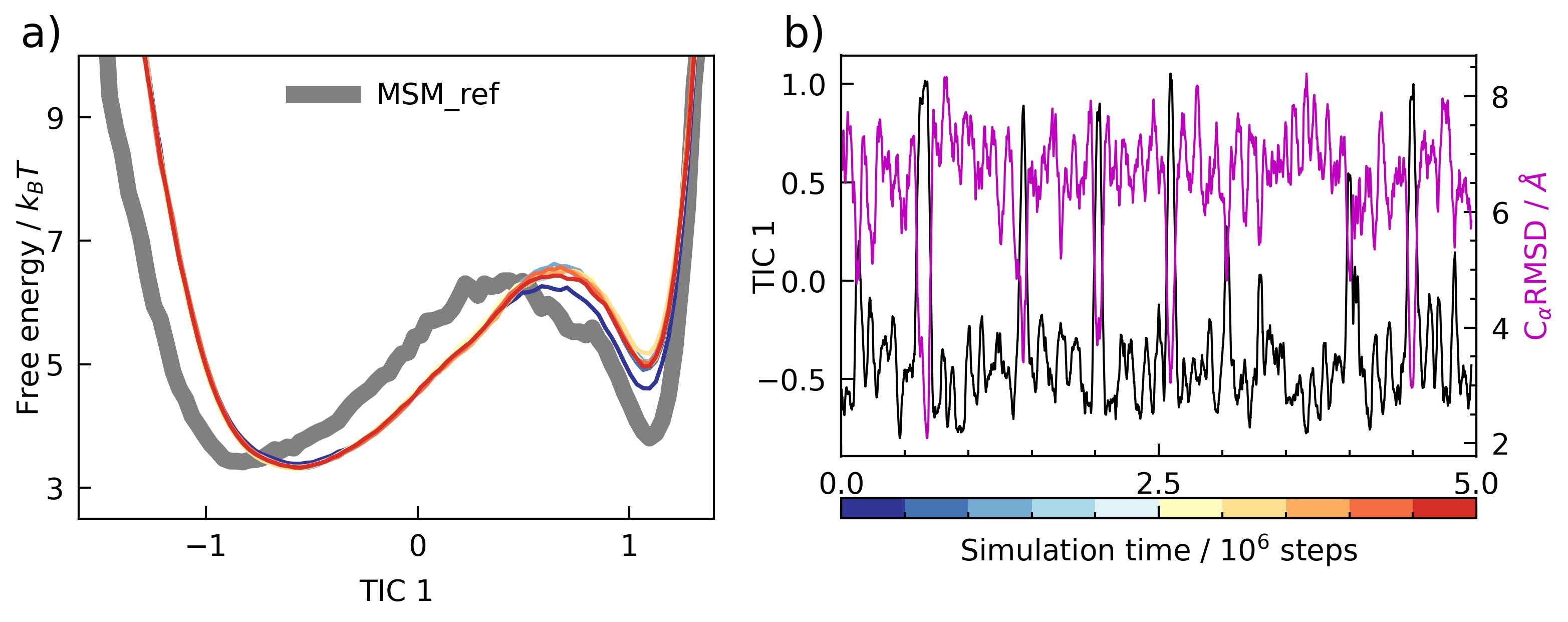}
\caption{Convergence of simulations for a Trp-Cage model.
a.~1-D histogram over the first TIC for models trained with a combination of denoising forces and atomistic forces on 2\% training set at noise level 0.003 $\text{\AA}^2$. Aggregated distribution from all 100 replicas in each of the 10 time windows is visualized with color reflecting the time frame.
The reference FES is shown in gray.
b.~1-D time series of a single trajectory
of the above simulation projected on the first TIC and RMSD.}
\label{fig:trpcage_convergence_appendix}
\end{figure*}

In order to demonstrate the convergence of the simulations, we split each trajectory into 10 non-overlapping chunks in time, each spanning 0.5M time steps. In Fig.~\ref{fig:trpcage_convergence_appendix}a an example is provided for noise level 0.003 \AA$^2$ trained on 2\% of available training data. We observed that after the initial shift of the distribution, the later chunks all lie within a small level of variation without systematic drift. 
There are frequent transitions between the folded and unfolded states in each trajectory 
(Figure~\ref{fig:trpcage_convergence_appendix}b). The C$_\alpha$-RMSD curve and the TIC curve outline the same trends, with short residence time in positive TIC~1 and small RMSD corresponding to the folded state. This coincides with the observation in Fig.~\ref{fig:trpcage_convergence_appendix}a and main text that the model is underestimating the stability of the folded state for this training set size and optimization procedure.

\begin{figure*}
\centering
\includegraphics[width=6.0in]{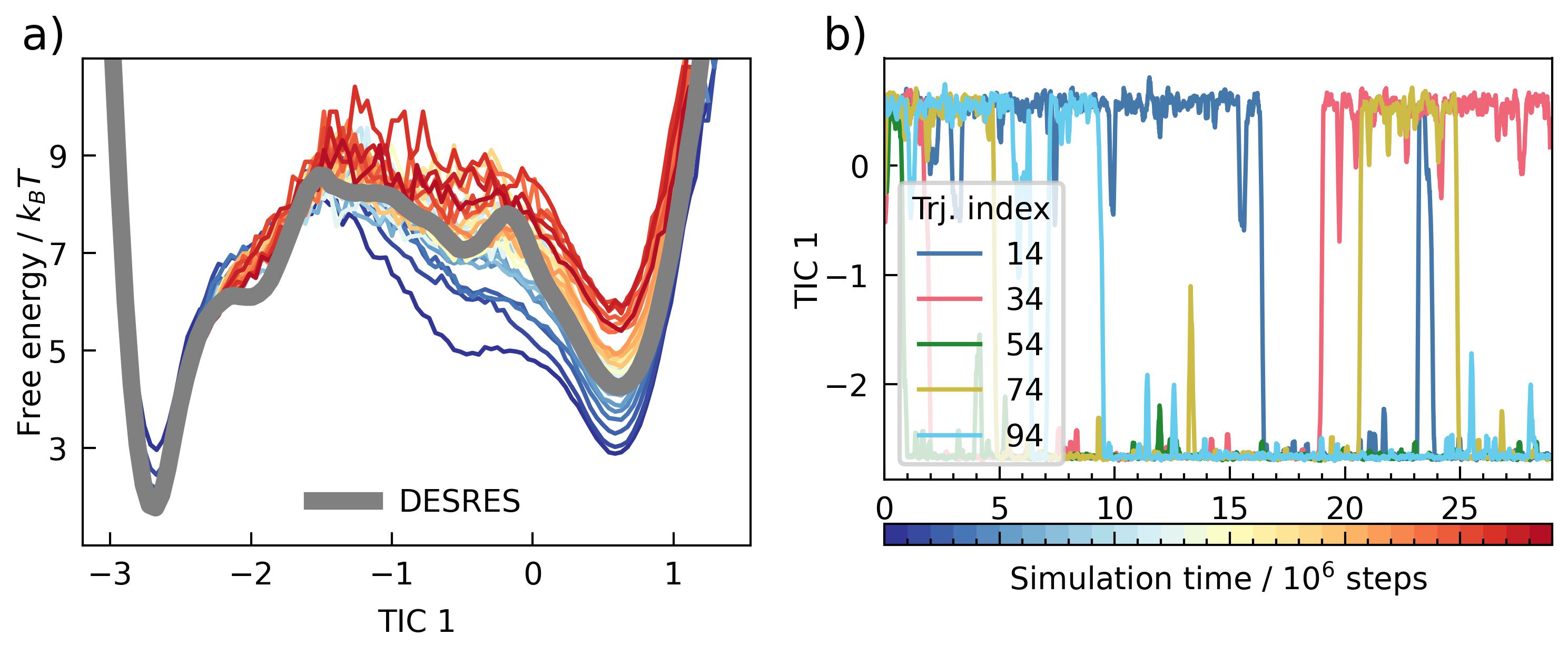}
\caption{Convergence of simulations for an NTL9 model.
a.~1-D histogram over the first TIC for models trained with a combination of denoising forces and atomistic forces on 2\% training set and noise level 0.003 $\text{\AA}^2$. Aggregated distribution from all 100 replicas in each time window is visualized according to the color bar.
The reference FES is shown in gray.
b.~1-D time series of five trajectories of the above simulation projected on the first TIC.}
\label{fig:ntl9_convergence_appendix}
\end{figure*}

\paragraph*{NTL9}
For NTL9 similar analyses were conducted in Fig.~\ref{fig:ntl9_convergence_appendix}. Owing to the slower kinetics, we observed a gradual shift of the distribution on TIC 1 over time, which reflects the evolution of the 100-replica ensemble from the biased starting distribution.
After a longer burn-in period (curves in blue, cyan and yellow), evolution along the FES converges, indicating that the simulations together are sampling the Boltzmann distribution.
Figure~\ref{fig:ntl9_convergence_appendix}b shows the TIC time series of 5 randomly picked trajectories, demonstrating bidirectional transitions between the folded and unfolded states. The RMSD curves are omitted for visual clarity.

\subsubsection{Two-dimensional FESs}
\paragraph*{Measure of PMF error}
PMF errors presented in the main text were calculated by creating histograms across the leading two TICs.
Bin definitions underlying the histograms were held constant when analyzing simulations of each molecule, and consisted of dividing each TIC axis into 100 equally sized windows between the maximum and minimum values observed across all corresponding models and reference simulations. The proportion of samples present in each window was transformed using $- \log $ and compared to reference values bin-wise using a square loss. This bin-wise loss was averaged using the population present in each bin as weights to create a single number characterizing accuracy. Bins without any samples were assigned a baseline proportion of $10^{-6}$. Trajectories were trimmed before error calculation using the same procedure as described below for the creation of 2D FES visualizations.

\begin{figure*}
\centering
\includegraphics[angle=90,origin=c,width=6.0in]{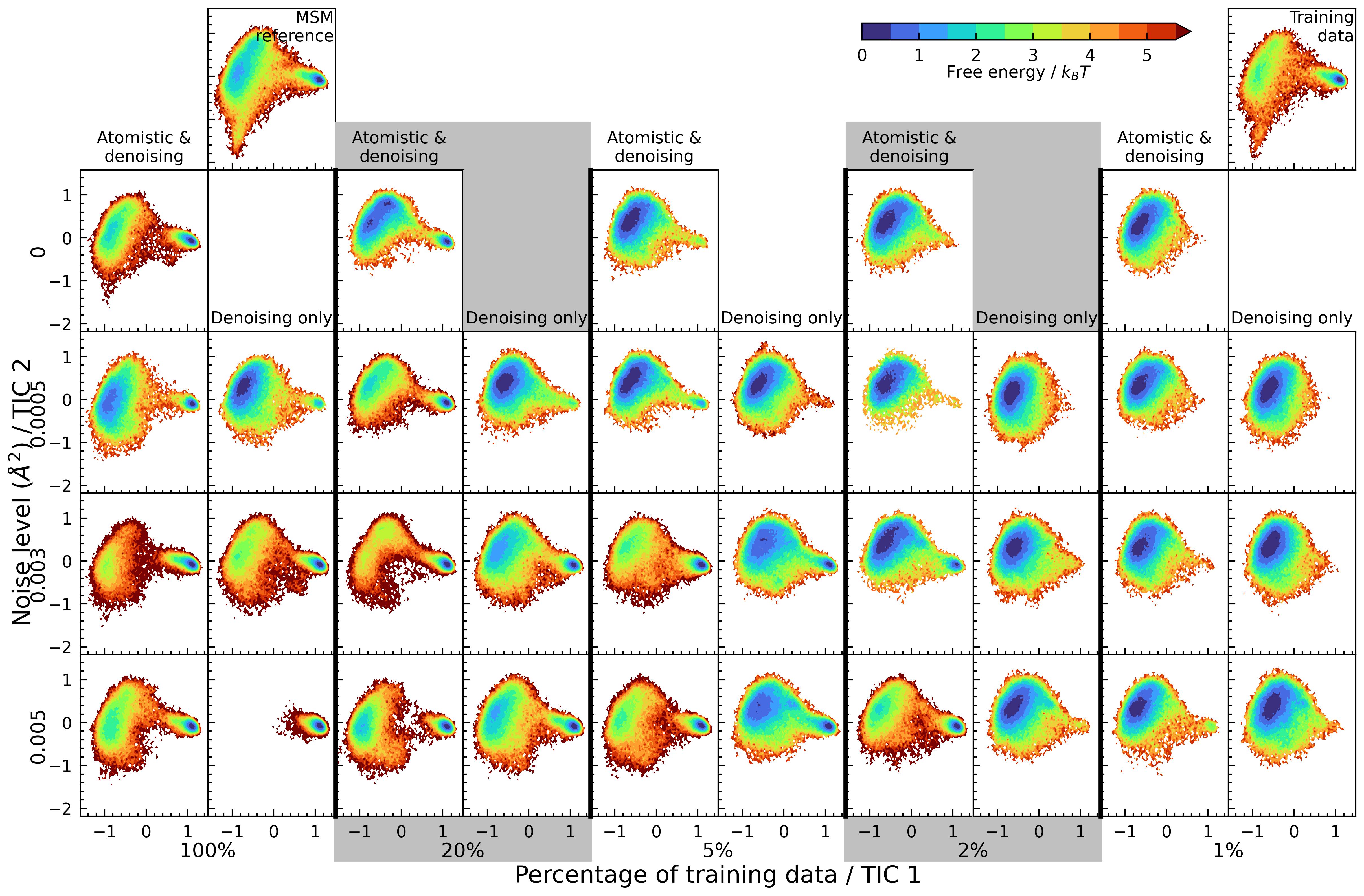}
\caption{2D FESs for Trp-Cage models.}
\label{fig:trpcage_2d_appendix}
\end{figure*}

\paragraph*{Trp-Cage} The FESs over the first and second TICs for all utilized models are visualized in Fig.~\ref{fig:trpcage_2d_appendix}.
As stated in the main text, models trained only on forces in the low-data regime cannot fold Trp-Cage, and the inclusion of noising
allows the model to correctly stabilize the folded state.
The second TIC resolves mainly the conformational kinetics in the unfolded state, which was preserved among all models that recovered the unfolded state (Fig.~\ref{fig:trpcage_2d_appendix}). 

\begin{figure*}
\centering
\includegraphics[angle=90,origin=c,width=6.0in]{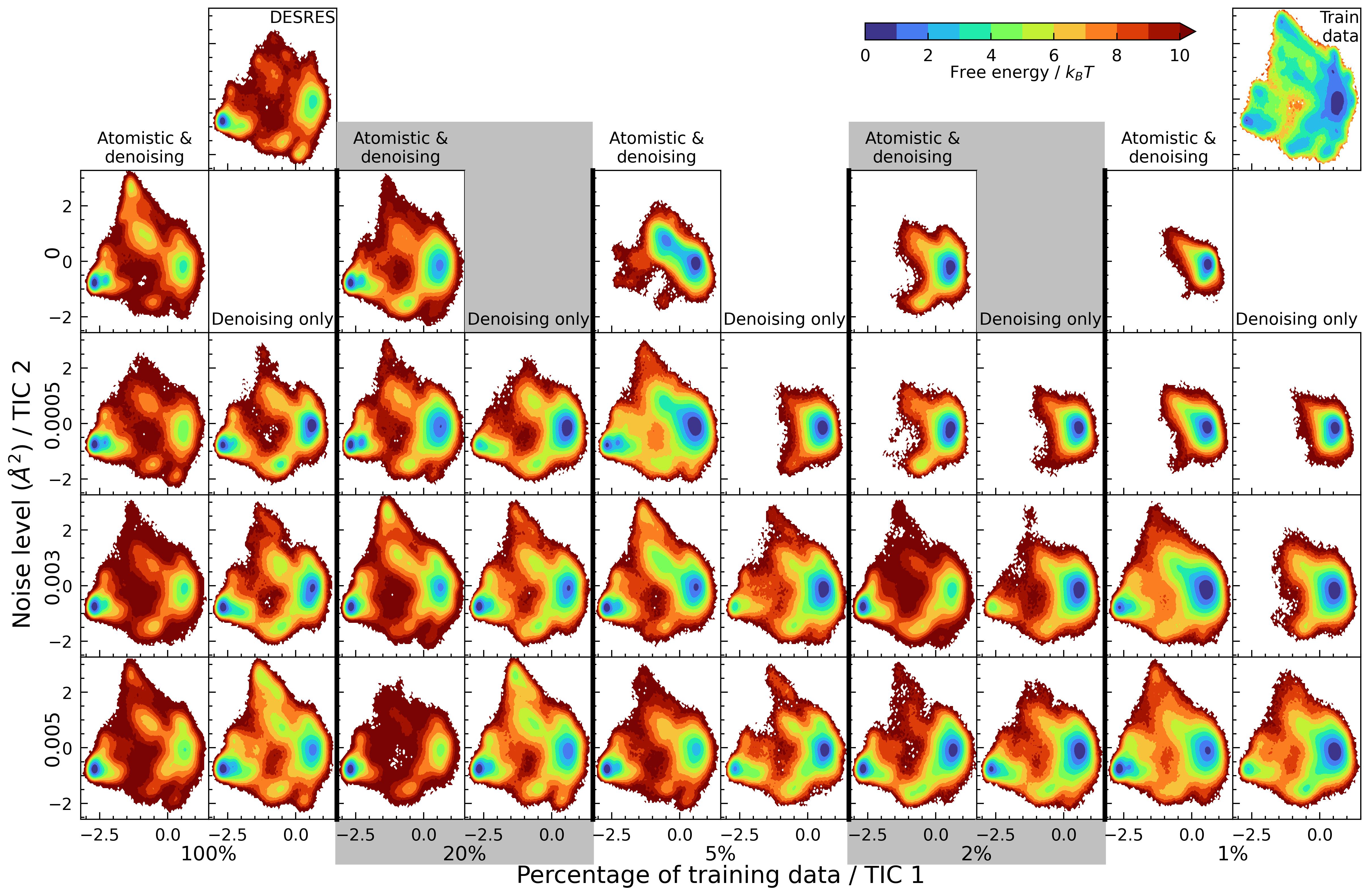}
\caption{2D FESs for NTL9 models.}
\label{fig:ntl9_2d_appendix}
\end{figure*}

\paragraph*{NTL9} Corresponding 2D FESs are found for NTL9 in Figure~\ref{fig:ntl9_2d_appendix}. Similar to the Trp-Cage, when focusing on the folded (bottom left) and unfolded (center right) states, the same trend as discussed in the main text is observed. Unlike in the case of Trp-Cage, the second TIC helps to distinguish several misfolded states and folding intermediates. Moreover, we observed that the distribution of those minor states does not converge to the reference ANTON simulations monotonically. We attribute this phenomenon to the fact that the transitions connecting those states are not thoroughly sampled in the atomistic reference despite the extensive simulations, as well as their slightly altered temperature. 
\clearpage

\bibliography{bibliography_sorted}

\begin{thebibliography}{81}%
\makeatletter
\providecommand \@ifxundefined [1]{%
 \@ifx{#1\undefined}
}%
\providecommand \@ifnum [1]{%
 \ifnum #1\expandafter \@firstoftwo
 \else \expandafter \@secondoftwo
 \fi
}%
\providecommand \@ifx [1]{%
 \ifx #1\expandafter \@firstoftwo
 \else \expandafter \@secondoftwo
 \fi
}%
\providecommand \natexlab [1]{#1}%
\providecommand \enquote  [1]{``#1''}%
\providecommand \bibnamefont  [1]{#1}%
\providecommand \bibfnamefont [1]{#1}%
\providecommand \citenamefont [1]{#1}%
\providecommand \href@noop [0]{\@secondoftwo}%
\providecommand \href [0]{\begingroup \@sanitize@url \@href}%
\providecommand \@href[1]{\@@startlink{#1}\@@href}%
\providecommand \@@href[1]{\endgroup#1\@@endlink}%
\providecommand \@sanitize@url [0]{\catcode `\\12\catcode `\$12\catcode
  `\&12\catcode `\#12\catcode `\^12\catcode `\_12\catcode `\%12\relax}%
\providecommand \@@startlink[1]{}%
\providecommand \@@endlink[0]{}%
\providecommand \url  [0]{\begingroup\@sanitize@url \@url }%
\providecommand \@url [1]{\endgroup\@href {#1}{\urlprefix }}%
\providecommand \urlprefix  [0]{URL }%
\providecommand \Eprint [0]{\href }%
\providecommand \doibase [0]{https://doi.org/}%
\providecommand \selectlanguage [0]{\@gobble}%
\providecommand \bibinfo  [0]{\@secondoftwo}%
\providecommand \bibfield  [0]{\@secondoftwo}%
\providecommand \translation [1]{[#1]}%
\providecommand \BibitemOpen [0]{}%
\providecommand \bibitemStop [0]{}%
\providecommand \bibitemNoStop [0]{.\EOS\space}%
\providecommand \EOS [0]{\spacefactor3000\relax}%
\providecommand \BibitemShut  [1]{\csname bibitem#1\endcsname}%
\let\auto@bib@innerbib\@empty
\bibitem [{\citenamefont {Adcock}\ and\ \citenamefont
  {McCammon}(2006)}]{adcock2006molecular}%
  \BibitemOpen
  \bibfield  {author} {\bibinfo {author} {\bibfnamefont {S.~A.}\ \bibnamefont
  {Adcock}}\ and\ \bibinfo {author} {\bibfnamefont {J.~A.}\ \bibnamefont
  {McCammon}},\ }\bibfield  {title} {\bibinfo {title} {Molecular dynamics:
  survey of methods for simulating the activity of proteins},\ }\href@noop {}
  {\bibfield  {journal} {\bibinfo  {journal} {Chem. Rev.}\ }\textbf {\bibinfo
  {volume} {106}},\ \bibinfo {pages} {1589} (\bibinfo {year}
  {2006})}\BibitemShut {NoStop}%
\bibitem [{\citenamefont {Hospital}\ \emph {et~al.}(2015)\citenamefont
  {Hospital}, \citenamefont {Go{\~n}i}, \citenamefont {Orozco},\ and\
  \citenamefont {Gelp{\'\i}}}]{hospital2015molecular}%
  \BibitemOpen
  \bibfield  {author} {\bibinfo {author} {\bibfnamefont {A.}~\bibnamefont
  {Hospital}}, \bibinfo {author} {\bibfnamefont {J.~R.}\ \bibnamefont
  {Go{\~n}i}}, \bibinfo {author} {\bibfnamefont {M.}~\bibnamefont {Orozco}},\
  and\ \bibinfo {author} {\bibfnamefont {J.~L.}\ \bibnamefont {Gelp{\'\i}}},\
  }\bibfield  {title} {\bibinfo {title} {Molecular dynamics simulations:
  advances and applications},\ }\href@noop {} {\bibfield  {journal} {\bibinfo
  {journal} {Adv. Appl. Bioinform. Chem.}\ ,\ \bibinfo {pages} {37}} (\bibinfo
  {year} {2015})}\BibitemShut {NoStop}%
\bibitem [{\citenamefont {Hollingsworth}\ and\ \citenamefont
  {Dror}(2018)}]{hollingsworth2018molecular}%
  \BibitemOpen
  \bibfield  {author} {\bibinfo {author} {\bibfnamefont {S.~A.}\ \bibnamefont
  {Hollingsworth}}\ and\ \bibinfo {author} {\bibfnamefont {R.~O.}\ \bibnamefont
  {Dror}},\ }\bibfield  {title} {\bibinfo {title} {Molecular dynamics
  simulation for all},\ }\href@noop {} {\bibfield  {journal} {\bibinfo
  {journal} {Neuron}\ }\textbf {\bibinfo {volume} {99}},\ \bibinfo {pages}
  {1129} (\bibinfo {year} {2018})}\BibitemShut {NoStop}%
\bibitem [{\citenamefont {Schlick}\ \emph {et~al.}(2021)\citenamefont
  {Schlick}, \citenamefont {Portillo-Ledesma}, \citenamefont {Myers},
  \citenamefont {Beljak}, \citenamefont {Chen}, \citenamefont {Dakhel},
  \citenamefont {Darling}, \citenamefont {Ghosh}, \citenamefont {Hall},
  \citenamefont {Jan} \emph {et~al.}}]{schlick2021biomolecular}%
  \BibitemOpen
  \bibfield  {author} {\bibinfo {author} {\bibfnamefont {T.}~\bibnamefont
  {Schlick}}, \bibinfo {author} {\bibfnamefont {S.}~\bibnamefont
  {Portillo-Ledesma}}, \bibinfo {author} {\bibfnamefont {C.~G.}\ \bibnamefont
  {Myers}}, \bibinfo {author} {\bibfnamefont {L.}~\bibnamefont {Beljak}},
  \bibinfo {author} {\bibfnamefont {J.}~\bibnamefont {Chen}}, \bibinfo {author}
  {\bibfnamefont {S.}~\bibnamefont {Dakhel}}, \bibinfo {author} {\bibfnamefont
  {D.}~\bibnamefont {Darling}}, \bibinfo {author} {\bibfnamefont
  {S.}~\bibnamefont {Ghosh}}, \bibinfo {author} {\bibfnamefont
  {J.}~\bibnamefont {Hall}}, \bibinfo {author} {\bibfnamefont {M.}~\bibnamefont
  {Jan}}, \emph {et~al.},\ }\bibfield  {title} {\bibinfo {title} {Biomolecular
  modeling and simulation: a prospering multidisciplinary field},\ }\href@noop
  {} {\bibfield  {journal} {\bibinfo  {journal} {Annu. Rev. Biophys.}\ }\textbf
  {\bibinfo {volume} {50}},\ \bibinfo {pages} {267} (\bibinfo {year}
  {2021})}\BibitemShut {NoStop}%
\bibitem [{\citenamefont {Clementi}(2008)}]{clementiCOSB2008}%
  \BibitemOpen
  \bibfield  {author} {\bibinfo {author} {\bibfnamefont {C.}~\bibnamefont
  {Clementi}},\ }\bibfield  {title} {\bibinfo {title} {Coarse-grained models of
  protein folding: toy models or predictive tools?},\ }\href@noop {} {\bibfield
   {journal} {\bibinfo  {journal} {Curr. Opin. Struct. Biol.}\ }\textbf
  {\bibinfo {volume} {18}},\ \bibinfo {pages} {10} (\bibinfo {year}
  {2008})}\BibitemShut {NoStop}%
\bibitem [{\citenamefont {Noid}(2013)}]{noid2013perspective}%
  \BibitemOpen
  \bibfield  {author} {\bibinfo {author} {\bibfnamefont {W.~G.}\ \bibnamefont
  {Noid}},\ }\bibfield  {title} {\bibinfo {title} {Perspective: Coarse-grained
  models for biomolecular systems},\ }\href@noop {} {\bibfield  {journal}
  {\bibinfo  {journal} {J. Chem. Phys.}\ }\textbf {\bibinfo {volume} {139}},\
  \bibinfo {pages} {090901} (\bibinfo {year} {2013})}\BibitemShut {NoStop}%
\bibitem [{\citenamefont {Liwo}\ \emph {et~al.}(2021)\citenamefont {Liwo},
  \citenamefont {Czaplewski}, \citenamefont {Sieradzan}, \citenamefont
  {Lipska}, \citenamefont {Samsonov},\ and\ \citenamefont
  {Murarka}}]{liwo2021theory}%
  \BibitemOpen
  \bibfield  {author} {\bibinfo {author} {\bibfnamefont {A.}~\bibnamefont
  {Liwo}}, \bibinfo {author} {\bibfnamefont {C.}~\bibnamefont {Czaplewski}},
  \bibinfo {author} {\bibfnamefont {A.~K.}\ \bibnamefont {Sieradzan}}, \bibinfo
  {author} {\bibfnamefont {A.~G.}\ \bibnamefont {Lipska}}, \bibinfo {author}
  {\bibfnamefont {S.~A.}\ \bibnamefont {Samsonov}},\ and\ \bibinfo {author}
  {\bibfnamefont {R.~K.}\ \bibnamefont {Murarka}},\ }\bibfield  {title}
  {\bibinfo {title} {Theory and practice of coarse-grained molecular dynamics
  of biologically important systems},\ }\href@noop {} {\bibfield  {journal}
  {\bibinfo  {journal} {Biomolecules}\ }\textbf {\bibinfo {volume} {11}},\
  \bibinfo {pages} {1347} (\bibinfo {year} {2021})}\BibitemShut {NoStop}%
\bibitem [{\citenamefont {Jin}\ \emph {et~al.}(2022)\citenamefont {Jin},
  \citenamefont {Pak}, \citenamefont {Durumeric}, \citenamefont {Loose},\ and\
  \citenamefont {Voth}}]{jin2022bottom}%
  \BibitemOpen
  \bibfield  {author} {\bibinfo {author} {\bibfnamefont {J.}~\bibnamefont
  {Jin}}, \bibinfo {author} {\bibfnamefont {A.~J.}\ \bibnamefont {Pak}},
  \bibinfo {author} {\bibfnamefont {A.~E.}\ \bibnamefont {Durumeric}}, \bibinfo
  {author} {\bibfnamefont {T.~D.}\ \bibnamefont {Loose}},\ and\ \bibinfo
  {author} {\bibfnamefont {G.~A.}\ \bibnamefont {Voth}},\ }\bibfield  {title}
  {\bibinfo {title} {Bottom-up coarse-graining: Principles and perspectives},\
  }\href@noop {} {\bibfield  {journal} {\bibinfo  {journal} {J. Chem. Theory
  Comput.}\ }\textbf {\bibinfo {volume} {18}},\ \bibinfo {pages} {5759}
  (\bibinfo {year} {2022})}\BibitemShut {NoStop}%
\bibitem [{\citenamefont {Noid}(2023)}]{noid2023perspective}%
  \BibitemOpen
  \bibfield  {author} {\bibinfo {author} {\bibfnamefont {W.~G.}\ \bibnamefont
  {Noid}},\ }\bibfield  {title} {\bibinfo {title} {Perspective: Advances,
  challenges, and insight for predictive coarse-grained models},\ }\href@noop
  {} {\bibfield  {journal} {\bibinfo  {journal} {J. Phys. Chem. B}\ }\textbf
  {\bibinfo {volume} {127}},\ \bibinfo {pages} {4174} (\bibinfo {year}
  {2023})}\BibitemShut {NoStop}%
\bibitem [{\citenamefont {Borges-Ara{\'u}jo}\ \emph {et~al.}(2023)\citenamefont
  {Borges-Ara{\'u}jo}, \citenamefont {Patmanidis}, \citenamefont {Singh},
  \citenamefont {Santos}, \citenamefont {Sieradzan}, \citenamefont {Vanni},
  \citenamefont {Czaplewski}, \citenamefont {Pantano}, \citenamefont {Shinoda},
  \citenamefont {Monticelli} \emph {et~al.}}]{borges2023pragmatic}%
  \BibitemOpen
  \bibfield  {author} {\bibinfo {author} {\bibfnamefont {L.}~\bibnamefont
  {Borges-Ara{\'u}jo}}, \bibinfo {author} {\bibfnamefont {I.}~\bibnamefont
  {Patmanidis}}, \bibinfo {author} {\bibfnamefont {A.~P.}\ \bibnamefont
  {Singh}}, \bibinfo {author} {\bibfnamefont {L.~H.}\ \bibnamefont {Santos}},
  \bibinfo {author} {\bibfnamefont {A.~K.}\ \bibnamefont {Sieradzan}}, \bibinfo
  {author} {\bibfnamefont {S.}~\bibnamefont {Vanni}}, \bibinfo {author}
  {\bibfnamefont {C.}~\bibnamefont {Czaplewski}}, \bibinfo {author}
  {\bibfnamefont {S.}~\bibnamefont {Pantano}}, \bibinfo {author} {\bibfnamefont
  {W.}~\bibnamefont {Shinoda}}, \bibinfo {author} {\bibfnamefont
  {L.}~\bibnamefont {Monticelli}}, \emph {et~al.},\ }\bibfield  {title}
  {\bibinfo {title} {Pragmatic coarse-graining of proteins: models and
  applications},\ }\href@noop {} {\bibfield  {journal} {\bibinfo  {journal} {J.
  Chem. Theory Comput.}\ }\textbf {\bibinfo {volume} {19}},\ \bibinfo {pages}
  {7112} (\bibinfo {year} {2023})}\BibitemShut {NoStop}%
\bibitem [{\citenamefont {Marrink}\ \emph {et~al.}(2023)\citenamefont
  {Marrink}, \citenamefont {Monticelli}, \citenamefont {Melo}, \citenamefont
  {Alessandri}, \citenamefont {Tieleman},\ and\ \citenamefont
  {Souza}}]{marrink2023two}%
  \BibitemOpen
  \bibfield  {author} {\bibinfo {author} {\bibfnamefont {S.~J.}\ \bibnamefont
  {Marrink}}, \bibinfo {author} {\bibfnamefont {L.}~\bibnamefont {Monticelli}},
  \bibinfo {author} {\bibfnamefont {M.~N.}\ \bibnamefont {Melo}}, \bibinfo
  {author} {\bibfnamefont {R.}~\bibnamefont {Alessandri}}, \bibinfo {author}
  {\bibfnamefont {D.~P.}\ \bibnamefont {Tieleman}},\ and\ \bibinfo {author}
  {\bibfnamefont {P.~C.}\ \bibnamefont {Souza}},\ }\bibfield  {title} {\bibinfo
  {title} {Two decades of martini: Better beads, broader scope},\ }\href@noop
  {} {\bibfield  {journal} {\bibinfo  {journal} {Wiley Interdisciplinary
  Reviews: Computational Molecular Science}\ }\textbf {\bibinfo {volume}
  {13}},\ \bibinfo {pages} {e1620} (\bibinfo {year} {2023})}\BibitemShut
  {NoStop}%
\bibitem [{\citenamefont {Robustelli}\ \emph {et~al.}(2018)\citenamefont
  {Robustelli}, \citenamefont {Piana},\ and\ \citenamefont
  {Shaw}}]{robustelli2018developing}%
  \BibitemOpen
  \bibfield  {author} {\bibinfo {author} {\bibfnamefont {P.}~\bibnamefont
  {Robustelli}}, \bibinfo {author} {\bibfnamefont {S.}~\bibnamefont {Piana}},\
  and\ \bibinfo {author} {\bibfnamefont {D.~E.}\ \bibnamefont {Shaw}},\
  }\bibfield  {title} {\bibinfo {title} {Developing a molecular dynamics force
  field for both folded and disordered protein states},\ }\href@noop {}
  {\bibfield  {journal} {\bibinfo  {journal} {Proc. Natl. Acad. Sci. U.S.A.}\
  }\textbf {\bibinfo {volume} {115}},\ \bibinfo {pages} {E4758} (\bibinfo
  {year} {2018})}\BibitemShut {NoStop}%
\bibitem [{\citenamefont {Best}(2019)}]{best2019atomistic}%
  \BibitemOpen
  \bibfield  {author} {\bibinfo {author} {\bibfnamefont {R.~B.}\ \bibnamefont
  {Best}},\ }\bibfield  {title} {\bibinfo {title} {Atomistic force fields for
  proteins},\ }\href@noop {} {\bibfield  {journal} {\bibinfo  {journal}
  {Biomolecular Simulations: Methods and Protocols}\ ,\ \bibinfo {pages} {3}}
  (\bibinfo {year} {2019})}\BibitemShut {NoStop}%
\bibitem [{\citenamefont {Shaw}\ \emph {et~al.}(2010)\citenamefont {Shaw},
  \citenamefont {Maragakis}, \citenamefont {Lindorff-Larsen}, \citenamefont
  {Piana}, \citenamefont {Dror}, \citenamefont {Eastwood}, \citenamefont
  {Bank}, \citenamefont {Jumper}, \citenamefont {Salmon}, \citenamefont
  {Shan},\ and\ \citenamefont {Wriggers}}]{shaw2010atomic}%
  \BibitemOpen
  \bibfield  {author} {\bibinfo {author} {\bibfnamefont {D.~E.}\ \bibnamefont
  {Shaw}}, \bibinfo {author} {\bibfnamefont {P.}~\bibnamefont {Maragakis}},
  \bibinfo {author} {\bibfnamefont {K.}~\bibnamefont {Lindorff-Larsen}},
  \bibinfo {author} {\bibfnamefont {S.}~\bibnamefont {Piana}}, \bibinfo
  {author} {\bibfnamefont {R.~O.}\ \bibnamefont {Dror}}, \bibinfo {author}
  {\bibfnamefont {M.~P.}\ \bibnamefont {Eastwood}}, \bibinfo {author}
  {\bibfnamefont {J.~A.}\ \bibnamefont {Bank}}, \bibinfo {author}
  {\bibfnamefont {J.~M.}\ \bibnamefont {Jumper}}, \bibinfo {author}
  {\bibfnamefont {J.~K.}\ \bibnamefont {Salmon}}, \bibinfo {author}
  {\bibfnamefont {Y.}~\bibnamefont {Shan}},\ and\ \bibinfo {author}
  {\bibfnamefont {W.}~\bibnamefont {Wriggers}},\ }\bibfield  {title} {\bibinfo
  {title} {Atomic-level characterization of the structural dynamics of
  proteins},\ }\href@noop {} {\bibfield  {journal} {\bibinfo  {journal}
  {Science}\ }\textbf {\bibinfo {volume} {330}},\ \bibinfo {pages} {341}
  (\bibinfo {year} {2010})}\BibitemShut {NoStop}%
\bibitem [{\citenamefont {Voelz}\ \emph {et~al.}(2010)\citenamefont {Voelz},
  \citenamefont {Bowman}, \citenamefont {Beauchamp},\ and\ \citenamefont
  {Pande}}]{voelz2010molecular}%
  \BibitemOpen
  \bibfield  {author} {\bibinfo {author} {\bibfnamefont {V.~A.}\ \bibnamefont
  {Voelz}}, \bibinfo {author} {\bibfnamefont {G.~R.}\ \bibnamefont {Bowman}},
  \bibinfo {author} {\bibfnamefont {K.}~\bibnamefont {Beauchamp}},\ and\
  \bibinfo {author} {\bibfnamefont {V.~S.}\ \bibnamefont {Pande}},\ }\bibfield
  {title} {\bibinfo {title} {Molecular simulation of ab initio protein folding
  for a millisecond folder ntl9 (1- 39)},\ }\href@noop {} {\bibfield  {journal}
  {\bibinfo  {journal} {J. Am. Chem. Soc.}\ }\textbf {\bibinfo {volume}
  {132}},\ \bibinfo {pages} {1526} (\bibinfo {year} {2010})}\BibitemShut
  {NoStop}%
\bibitem [{\citenamefont {Lindorff-Larsen}\ \emph {et~al.}(2011)\citenamefont
  {Lindorff-Larsen}, \citenamefont {Piana}, \citenamefont {Dror},\ and\
  \citenamefont {Shaw}}]{lindorff2011fast}%
  \BibitemOpen
  \bibfield  {author} {\bibinfo {author} {\bibfnamefont {K.}~\bibnamefont
  {Lindorff-Larsen}}, \bibinfo {author} {\bibfnamefont {S.}~\bibnamefont
  {Piana}}, \bibinfo {author} {\bibfnamefont {R.~O.}\ \bibnamefont {Dror}},\
  and\ \bibinfo {author} {\bibfnamefont {D.~E.}\ \bibnamefont {Shaw}},\
  }\bibfield  {title} {\bibinfo {title} {How fast-folding proteins fold},\
  }\href@noop {} {\bibfield  {journal} {\bibinfo  {journal} {Science}\ }\textbf
  {\bibinfo {volume} {334}},\ \bibinfo {pages} {517} (\bibinfo {year}
  {2011})}\BibitemShut {NoStop}%
\bibitem [{\citenamefont {Plattner}\ \emph {et~al.}(2017)\citenamefont
  {Plattner}, \citenamefont {Doerr}, \citenamefont {De~Fabritiis},\ and\
  \citenamefont {No{\'e}}}]{plattner2017complete}%
  \BibitemOpen
  \bibfield  {author} {\bibinfo {author} {\bibfnamefont {N.}~\bibnamefont
  {Plattner}}, \bibinfo {author} {\bibfnamefont {S.}~\bibnamefont {Doerr}},
  \bibinfo {author} {\bibfnamefont {G.}~\bibnamefont {De~Fabritiis}},\ and\
  \bibinfo {author} {\bibfnamefont {F.}~\bibnamefont {No{\'e}}},\ }\bibfield
  {title} {\bibinfo {title} {Complete protein--protein association kinetics in
  atomic detail revealed by molecular dynamics simulations and markov
  modelling},\ }\href@noop {} {\bibfield  {journal} {\bibinfo  {journal} {Nat.
  Chem.}\ }\textbf {\bibinfo {volume} {9}},\ \bibinfo {pages} {1005} (\bibinfo
  {year} {2017})}\BibitemShut {NoStop}%
\bibitem [{\citenamefont {Robustelli}\ \emph {et~al.}(2022)\citenamefont
  {Robustelli}, \citenamefont {Ibanez-de Opakua}, \citenamefont
  {Campbell-Bezat}, \citenamefont {Giordanetto}, \citenamefont {Becker},
  \citenamefont {Zweckstetter}, \citenamefont {Pan},\ and\ \citenamefont
  {Shaw}}]{robustelli2022molecular}%
  \BibitemOpen
  \bibfield  {author} {\bibinfo {author} {\bibfnamefont {P.}~\bibnamefont
  {Robustelli}}, \bibinfo {author} {\bibfnamefont {A.}~\bibnamefont {Ibanez-de
  Opakua}}, \bibinfo {author} {\bibfnamefont {C.}~\bibnamefont
  {Campbell-Bezat}}, \bibinfo {author} {\bibfnamefont {F.}~\bibnamefont
  {Giordanetto}}, \bibinfo {author} {\bibfnamefont {S.}~\bibnamefont {Becker}},
  \bibinfo {author} {\bibfnamefont {M.}~\bibnamefont {Zweckstetter}}, \bibinfo
  {author} {\bibfnamefont {A.~C.}\ \bibnamefont {Pan}},\ and\ \bibinfo {author}
  {\bibfnamefont {D.~E.}\ \bibnamefont {Shaw}},\ }\bibfield  {title} {\bibinfo
  {title} {Molecular basis of small-molecule binding to $\alpha$-synuclein},\
  }\href@noop {} {\bibfield  {journal} {\bibinfo  {journal} {J. Am. Chem.
  Soc.}\ }\textbf {\bibinfo {volume} {144}},\ \bibinfo {pages} {2501} (\bibinfo
  {year} {2022})}\BibitemShut {NoStop}%
\bibitem [{\citenamefont {Wang}\ \emph {et~al.}(2019)\citenamefont {Wang},
  \citenamefont {Olsson}, \citenamefont {Wehmeyer}, \citenamefont {P{\'e}rez},
  \citenamefont {Charron}, \citenamefont {Fabritiis}, \citenamefont {No{\'e}},\
  and\ \citenamefont {Clementi}}]{Wang2019}%
  \BibitemOpen
  \bibfield  {author} {\bibinfo {author} {\bibfnamefont {J.}~\bibnamefont
  {Wang}}, \bibinfo {author} {\bibfnamefont {S.}~\bibnamefont {Olsson}},
  \bibinfo {author} {\bibfnamefont {C.}~\bibnamefont {Wehmeyer}}, \bibinfo
  {author} {\bibfnamefont {A.}~\bibnamefont {P{\'e}rez}}, \bibinfo {author}
  {\bibfnamefont {N.~E.}\ \bibnamefont {Charron}}, \bibinfo {author}
  {\bibfnamefont {G.~D.}\ \bibnamefont {Fabritiis}}, \bibinfo {author}
  {\bibfnamefont {F.}~\bibnamefont {No{\'e}}},\ and\ \bibinfo {author}
  {\bibfnamefont {C.}~\bibnamefont {Clementi}},\ }\bibfield  {title} {\bibinfo
  {title} {Machine learning of coarse-grained molecular dynamics force
  fields},\ }\href {https://doi.org/10.1021/acscentsci.8b00913} {\bibfield
  {journal} {\bibinfo  {journal} {ACS Cent. Sci.}\ }\textbf {\bibinfo {volume}
  {5}},\ \bibinfo {pages} {755} (\bibinfo {year} {2019})}\BibitemShut {NoStop}%
\bibitem [{\citenamefont {Durumeric}\ \emph {et~al.}(2023)\citenamefont
  {Durumeric}, \citenamefont {Charron}, \citenamefont {Templeton},
  \citenamefont {Musil}, \citenamefont {Bonneau}, \citenamefont {Pasos-Trejo},
  \citenamefont {Chen}, \citenamefont {Kelkar}, \citenamefont {No{\'e}},\ and\
  \citenamefont {Clementi}}]{durumeric2023machine}%
  \BibitemOpen
  \bibfield  {author} {\bibinfo {author} {\bibfnamefont {A.~E.~P.}\
  \bibnamefont {Durumeric}}, \bibinfo {author} {\bibfnamefont {N.~E.}\
  \bibnamefont {Charron}}, \bibinfo {author} {\bibfnamefont {C.}~\bibnamefont
  {Templeton}}, \bibinfo {author} {\bibfnamefont {F.}~\bibnamefont {Musil}},
  \bibinfo {author} {\bibfnamefont {K.}~\bibnamefont {Bonneau}}, \bibinfo
  {author} {\bibfnamefont {A.~S.}\ \bibnamefont {Pasos-Trejo}}, \bibinfo
  {author} {\bibfnamefont {Y.}~\bibnamefont {Chen}}, \bibinfo {author}
  {\bibfnamefont {A.}~\bibnamefont {Kelkar}}, \bibinfo {author} {\bibfnamefont
  {F.}~\bibnamefont {No{\'e}}},\ and\ \bibinfo {author} {\bibfnamefont
  {C.}~\bibnamefont {Clementi}},\ }\bibfield  {title} {\bibinfo {title}
  {Machine learned coarse-grained protein force-fields: Are we there yet?},\
  }\href@noop {} {\bibfield  {journal} {\bibinfo  {journal} {Curr. Opin.
  Struct. Biol.}\ }\textbf {\bibinfo {volume} {79}},\ \bibinfo {pages} {102533}
  (\bibinfo {year} {2023})}\BibitemShut {NoStop}%
\bibitem [{\citenamefont {Majewski}\ \emph {et~al.}(2023)\citenamefont
  {Majewski}, \citenamefont {P{\'e}rez}, \citenamefont {Th\"{o}lke},
  \citenamefont {Doerr}, \citenamefont {Charron}, \citenamefont {Giorgino},
  \citenamefont {Husic}, \citenamefont {Clementi}, \citenamefont {No{\'e}},\
  and\ \citenamefont {De~Fabritiis}}]{majewski2022machine}%
  \BibitemOpen
  \bibfield  {author} {\bibinfo {author} {\bibfnamefont {M.}~\bibnamefont
  {Majewski}}, \bibinfo {author} {\bibfnamefont {A.}~\bibnamefont {P{\'e}rez}},
  \bibinfo {author} {\bibfnamefont {P.}~\bibnamefont {Th\"{o}lke}}, \bibinfo
  {author} {\bibfnamefont {S.}~\bibnamefont {Doerr}}, \bibinfo {author}
  {\bibfnamefont {N.~E.}\ \bibnamefont {Charron}}, \bibinfo {author}
  {\bibfnamefont {T.}~\bibnamefont {Giorgino}}, \bibinfo {author}
  {\bibfnamefont {B.~E.}\ \bibnamefont {Husic}}, \bibinfo {author}
  {\bibfnamefont {C.}~\bibnamefont {Clementi}}, \bibinfo {author}
  {\bibfnamefont {F.}~\bibnamefont {No{\'e}}},\ and\ \bibinfo {author}
  {\bibfnamefont {G.}~\bibnamefont {De~Fabritiis}},\ }\bibfield  {title}
  {\bibinfo {title} {Machine learning coarse-grained potentials of protein
  thermodynamics},\ }\href {http://dx.doi.org/10.1038/s41467-023-41343-1}
  {\bibfield  {journal} {\bibinfo  {journal} {Nat. Commun.}\ }\textbf {\bibinfo
  {volume} {14}} (\bibinfo {year} {2023})}\BibitemShut {NoStop}%
\bibitem [{\citenamefont {Charron}\ \emph {et~al.}(2023)\citenamefont
  {Charron}, \citenamefont {Musil}, \citenamefont {Guljas}, \citenamefont
  {Chen}, \citenamefont {Bonneau}, \citenamefont {Pasos-Trejo}, \citenamefont
  {Venturin}, \citenamefont {Gusew}, \citenamefont {Zaporozhets}, \citenamefont
  {Kr{\"a}mer}, \citenamefont {Templeton}, \citenamefont {Kelkar},
  \citenamefont {Durumeric}, \citenamefont {Olsson}, \citenamefont {P{\'e}rez},
  \citenamefont {Mejewski}, \citenamefont {Husic}, \citenamefont {Patel},
  \citenamefont {Gianni}, \citenamefont {No{\'e}},\ and\ \citenamefont
  {Clementi}}]{charron2023navigating}%
  \BibitemOpen
  \bibfield  {author} {\bibinfo {author} {\bibfnamefont {N.~E.}\ \bibnamefont
  {Charron}}, \bibinfo {author} {\bibfnamefont {F.}~\bibnamefont {Musil}},
  \bibinfo {author} {\bibfnamefont {A.}~\bibnamefont {Guljas}}, \bibinfo
  {author} {\bibfnamefont {Y.}~\bibnamefont {Chen}}, \bibinfo {author}
  {\bibfnamefont {K.}~\bibnamefont {Bonneau}}, \bibinfo {author} {\bibfnamefont
  {A.~S.}\ \bibnamefont {Pasos-Trejo}}, \bibinfo {author} {\bibfnamefont
  {J.}~\bibnamefont {Venturin}}, \bibinfo {author} {\bibfnamefont
  {D.}~\bibnamefont {Gusew}}, \bibinfo {author} {\bibfnamefont
  {I.}~\bibnamefont {Zaporozhets}}, \bibinfo {author} {\bibfnamefont
  {A.}~\bibnamefont {Kr{\"a}mer}}, \bibinfo {author} {\bibfnamefont
  {C.}~\bibnamefont {Templeton}}, \bibinfo {author} {\bibfnamefont
  {A.}~\bibnamefont {Kelkar}}, \bibinfo {author} {\bibfnamefont {A.~E.~P.}\
  \bibnamefont {Durumeric}}, \bibinfo {author} {\bibfnamefont {S.}~\bibnamefont
  {Olsson}}, \bibinfo {author} {\bibfnamefont {A.}~\bibnamefont {P{\'e}rez}},
  \bibinfo {author} {\bibfnamefont {M.}~\bibnamefont {Mejewski}}, \bibinfo
  {author} {\bibfnamefont {B.~E.}\ \bibnamefont {Husic}}, \bibinfo {author}
  {\bibfnamefont {A.}~\bibnamefont {Patel}}, \bibinfo {author} {\bibfnamefont
  {F.~D.}\ \bibnamefont {Gianni}}, \bibinfo {author} {\bibfnamefont
  {F.}~\bibnamefont {No{\'e}}},\ and\ \bibinfo {author} {\bibfnamefont
  {C.}~\bibnamefont {Clementi}},\ }\bibfield  {title} {\bibinfo {title}
  {Navigating protein landscapes with a machine-learned transferable
  coarse-grained model},\ }\href@noop {} {\bibfield  {journal} {\bibinfo
  {journal} {arXiv preprint arXiv:2310.18278}\ } (\bibinfo {year}
  {2023})}\BibitemShut {NoStop}%
\bibitem [{\citenamefont {Wang}\ \emph {et~al.}(2021)\citenamefont {Wang},
  \citenamefont {Charron}, \citenamefont {Husic}, \citenamefont {Olsson},
  \citenamefont {No{\'e}},\ and\ \citenamefont {Clementi}}]{wang2021multibody}%
  \BibitemOpen
  \bibfield  {author} {\bibinfo {author} {\bibfnamefont {J.}~\bibnamefont
  {Wang}}, \bibinfo {author} {\bibfnamefont {N.}~\bibnamefont {Charron}},
  \bibinfo {author} {\bibfnamefont {B.}~\bibnamefont {Husic}}, \bibinfo
  {author} {\bibfnamefont {S.}~\bibnamefont {Olsson}}, \bibinfo {author}
  {\bibfnamefont {F.}~\bibnamefont {No{\'e}}},\ and\ \bibinfo {author}
  {\bibfnamefont {C.}~\bibnamefont {Clementi}},\ }\bibfield  {title} {\bibinfo
  {title} {Multi-body effects in a coarse-grained protein force field},\ }\href
  {https://doi.org/10.1063/5.0041022} {\bibfield  {journal} {\bibinfo
  {journal} {J. Chem. Phys.}\ }\textbf {\bibinfo {volume} {154}},\ \bibinfo
  {pages} {164113} (\bibinfo {year} {2021})}\BibitemShut {NoStop}%
\bibitem [{\citenamefont {Zaporozhets}\ and\ \citenamefont
  {Clementi}(2023)}]{zaporozhets2023multibody}%
  \BibitemOpen
  \bibfield  {author} {\bibinfo {author} {\bibfnamefont {I.}~\bibnamefont
  {Zaporozhets}}\ and\ \bibinfo {author} {\bibfnamefont {C.}~\bibnamefont
  {Clementi}},\ }\bibfield  {title} {\bibinfo {title} {Multibody terms in
  protein coarse-grained models: A top-down perspective},\ }\href@noop {}
  {\bibfield  {journal} {\bibinfo  {journal} {J. Phys. Chem. B}\ }\textbf
  {\bibinfo {volume} {127}},\ \bibinfo {pages} {6920} (\bibinfo {year}
  {2023})}\BibitemShut {NoStop}%
\bibitem [{\citenamefont {Izvekov}\ and\ \citenamefont
  {Voth}(2005{\natexlab{a}})}]{izvekov2005multiscale}%
  \BibitemOpen
  \bibfield  {author} {\bibinfo {author} {\bibfnamefont {S.}~\bibnamefont
  {Izvekov}}\ and\ \bibinfo {author} {\bibfnamefont {G.~A.}\ \bibnamefont
  {Voth}},\ }\bibfield  {title} {\bibinfo {title} {A multiscale coarse-graining
  method for biomolecular systems},\ }\href@noop {} {\bibfield  {journal}
  {\bibinfo  {journal} {J. Phys. Chem. B}\ }\textbf {\bibinfo {volume} {109}},\
  \bibinfo {pages} {2469} (\bibinfo {year} {2005}{\natexlab{a}})}\BibitemShut
  {NoStop}%
\bibitem [{\citenamefont {Izvekov}\ and\ \citenamefont
  {Voth}(2005{\natexlab{b}})}]{izvekov2005liquidmultiscale}%
  \BibitemOpen
  \bibfield  {author} {\bibinfo {author} {\bibfnamefont {S.}~\bibnamefont
  {Izvekov}}\ and\ \bibinfo {author} {\bibfnamefont {G.~A.}\ \bibnamefont
  {Voth}},\ }\bibfield  {title} {\bibinfo {title} {Multiscale coarse graining
  of liquid-state systems},\ }\href@noop {} {\bibfield  {journal} {\bibinfo
  {journal} {J. Chem. Phys.}\ }\textbf {\bibinfo {volume} {123}} (\bibinfo
  {year} {2005}{\natexlab{b}})}\BibitemShut {NoStop}%
\bibitem [{\citenamefont {Noid}\ \emph {et~al.}(2008)\citenamefont {Noid},
  \citenamefont {Chu}, \citenamefont {Ayton}, \citenamefont {Krishna},
  \citenamefont {Izvekov}, \citenamefont {Voth}, \citenamefont {Das},\ and\
  \citenamefont {Andersen}}]{Noid2008}%
  \BibitemOpen
  \bibfield  {author} {\bibinfo {author} {\bibfnamefont {W.~G.}\ \bibnamefont
  {Noid}}, \bibinfo {author} {\bibfnamefont {J.~W.}\ \bibnamefont {Chu}},
  \bibinfo {author} {\bibfnamefont {G.~S.}\ \bibnamefont {Ayton}}, \bibinfo
  {author} {\bibfnamefont {V.}~\bibnamefont {Krishna}}, \bibinfo {author}
  {\bibfnamefont {S.}~\bibnamefont {Izvekov}}, \bibinfo {author} {\bibfnamefont
  {G.~A.}\ \bibnamefont {Voth}}, \bibinfo {author} {\bibfnamefont
  {A.}~\bibnamefont {Das}},\ and\ \bibinfo {author} {\bibfnamefont {H.~C.}\
  \bibnamefont {Andersen}},\ }\bibfield  {title} {\bibinfo {title} {The
  multiscale coarse-graining method. i. a rigorous bridge between atomistic and
  coarse-grained models},\ }\href {https://doi.org/10.1063/1.2938860}
  {\bibfield  {journal} {\bibinfo  {journal} {J. Chem. Phys.}\ }\textbf
  {\bibinfo {volume} {128}},\ \bibinfo {pages} {244114} (\bibinfo {year}
  {2008})}\BibitemShut {NoStop}%
\bibitem [{\citenamefont {Lu}\ \emph {et~al.}(2010)\citenamefont {Lu},
  \citenamefont {Izvekov}, \citenamefont {Das}, \citenamefont {Andersen},\ and\
  \citenamefont {Voth}}]{lu2010efficient}%
  \BibitemOpen
  \bibfield  {author} {\bibinfo {author} {\bibfnamefont {L.}~\bibnamefont
  {Lu}}, \bibinfo {author} {\bibfnamefont {S.}~\bibnamefont {Izvekov}},
  \bibinfo {author} {\bibfnamefont {A.}~\bibnamefont {Das}}, \bibinfo {author}
  {\bibfnamefont {H.~C.}\ \bibnamefont {Andersen}},\ and\ \bibinfo {author}
  {\bibfnamefont {G.~A.}\ \bibnamefont {Voth}},\ }\bibfield  {title} {\bibinfo
  {title} {Efficient, regularized, and scalable algorithms for multiscale
  coarse-graining},\ }\href@noop {} {\bibfield  {journal} {\bibinfo  {journal}
  {J. Chem. Theory Comput.}\ }\textbf {\bibinfo {volume} {6}},\ \bibinfo
  {pages} {954} (\bibinfo {year} {2010})}\BibitemShut {NoStop}%
\bibitem [{\citenamefont {Lu}\ and\ \citenamefont
  {Voth}(2012)}]{lu2012multiscale}%
  \BibitemOpen
  \bibfield  {author} {\bibinfo {author} {\bibfnamefont {L.}~\bibnamefont
  {Lu}}\ and\ \bibinfo {author} {\bibfnamefont {G.~A.}\ \bibnamefont {Voth}},\
  }\bibfield  {title} {\bibinfo {title} {The multiscale coarse-graining
  method},\ }\href@noop {} {\bibfield  {journal} {\bibinfo  {journal} {Adv.
  Chem. Phys.}\ }\textbf {\bibinfo {volume} {149}},\ \bibinfo {pages} {47}
  (\bibinfo {year} {2012})}\BibitemShut {NoStop}%
\bibitem [{\citenamefont {Bishop}\ and\ \citenamefont
  {Bishop}(2024)}]{bishop2024deep}%
  \BibitemOpen
  \bibfield  {author} {\bibinfo {author} {\bibfnamefont {C.~M.}\ \bibnamefont
  {Bishop}}\ and\ \bibinfo {author} {\bibfnamefont {H.}~\bibnamefont
  {Bishop}},\ }\href@noop {} {\emph {\bibinfo {title} {Deep learning:
  Foundations and concepts}}},\ Vol.~\bibinfo {volume} {1}\ (\bibinfo
  {publisher} {Springer},\ \bibinfo {year} {2024})\BibitemShut {NoStop}%
\bibitem [{\citenamefont {Lemke}\ and\ \citenamefont
  {Peter}(2017)}]{lemke2017neural}%
  \BibitemOpen
  \bibfield  {author} {\bibinfo {author} {\bibfnamefont {T.}~\bibnamefont
  {Lemke}}\ and\ \bibinfo {author} {\bibfnamefont {C.}~\bibnamefont {Peter}},\
  }\bibfield  {title} {\bibinfo {title} {Neural network based prediction of
  conformational free energies-a new route toward coarse-grained simulation
  models},\ }\href@noop {} {\bibfield  {journal} {\bibinfo  {journal} {J. Chem.
  Theory Comput.}\ }\textbf {\bibinfo {volume} {13}},\ \bibinfo {pages} {6213}
  (\bibinfo {year} {2017})}\BibitemShut {NoStop}%
\bibitem [{\citenamefont {Husic}\ \emph {et~al.}(2020)\citenamefont {Husic},
  \citenamefont {Charron}, \citenamefont {Lemm}, \citenamefont {Wang},
  \citenamefont {P{\'e}rez}, \citenamefont {Majewski}, \citenamefont
  {Kr{\"a}mer}, \citenamefont {Chen}, \citenamefont {Olsson}, \citenamefont
  {Fabritiis}, \citenamefont {No{\'e}},\ and\ \citenamefont
  {Clementi}}]{Husic2020}%
  \BibitemOpen
  \bibfield  {author} {\bibinfo {author} {\bibfnamefont {B.~E.}\ \bibnamefont
  {Husic}}, \bibinfo {author} {\bibfnamefont {N.~E.}\ \bibnamefont {Charron}},
  \bibinfo {author} {\bibfnamefont {D.}~\bibnamefont {Lemm}}, \bibinfo {author}
  {\bibfnamefont {J.}~\bibnamefont {Wang}}, \bibinfo {author} {\bibfnamefont
  {A.}~\bibnamefont {P{\'e}rez}}, \bibinfo {author} {\bibfnamefont
  {M.}~\bibnamefont {Majewski}}, \bibinfo {author} {\bibfnamefont
  {A.}~\bibnamefont {Kr{\"a}mer}}, \bibinfo {author} {\bibfnamefont
  {Y.}~\bibnamefont {Chen}}, \bibinfo {author} {\bibfnamefont {S.}~\bibnamefont
  {Olsson}}, \bibinfo {author} {\bibfnamefont {G.~D.}\ \bibnamefont
  {Fabritiis}}, \bibinfo {author} {\bibfnamefont {F.}~\bibnamefont {No{\'e}}},\
  and\ \bibinfo {author} {\bibfnamefont {C.}~\bibnamefont {Clementi}},\
  }\bibfield  {title} {\bibinfo {title} {Coarse graining molecular dynamics
  with graph neural networks},\ }\href {https://doi.org/10.1063/5.0026133}
  {\bibfield  {journal} {\bibinfo  {journal} {J. Chem. Phys.}\ }\textbf
  {\bibinfo {volume} {153}},\ \bibinfo {pages} {194101} (\bibinfo {year}
  {2020})}\BibitemShut {NoStop}%
\bibitem [{\citenamefont {Ding}\ and\ \citenamefont
  {Zhang}(2022)}]{Ding2022coarsegrained}%
  \BibitemOpen
  \bibfield  {author} {\bibinfo {author} {\bibfnamefont {X.}~\bibnamefont
  {Ding}}\ and\ \bibinfo {author} {\bibfnamefont {B.}~\bibnamefont {Zhang}},\
  }\bibfield  {title} {\bibinfo {title} {Contrastive learning of coarse-grained
  force fields},\ }\href
  {https://doi.org/10.1021/ACS.JCTC.2C00616/ASSET/IMAGES/LARGE/CT2C00616_0005.JPEG}
  {\bibfield  {journal} {\bibinfo  {journal} {J. Chem. Theory Comput.}\
  }\textbf {\bibinfo {volume} {18}},\ \bibinfo {pages} {6334} (\bibinfo {year}
  {2022})}\BibitemShut {NoStop}%
\bibitem [{\citenamefont {K{\"o}hler}\ \emph {et~al.}(2023)\citenamefont
  {K{\"o}hler}, \citenamefont {Chen}, \citenamefont {Kr{\"a}mer}, \citenamefont
  {Clementi},\ and\ \citenamefont {No{\'e}}}]{flowmatching2023}%
  \BibitemOpen
  \bibfield  {author} {\bibinfo {author} {\bibfnamefont {J.}~\bibnamefont
  {K{\"o}hler}}, \bibinfo {author} {\bibfnamefont {Y.}~\bibnamefont {Chen}},
  \bibinfo {author} {\bibfnamefont {A.}~\bibnamefont {Kr{\"a}mer}}, \bibinfo
  {author} {\bibfnamefont {C.}~\bibnamefont {Clementi}},\ and\ \bibinfo
  {author} {\bibfnamefont {F.}~\bibnamefont {No{\'e}}},\ }\bibfield  {title}
  {\bibinfo {title} {Flow-matching: Efficient coarse-graining of molecular
  dynamics without forces},\ }\href {https://doi.org/10.1021/ACS.JCTC.3C00016}
  {\bibfield  {journal} {\bibinfo  {journal} {J. Chem. Theory Comput.}\
  }\textbf {\bibinfo {volume} {19}},\ \bibinfo {pages} {942} (\bibinfo {year}
  {2023})}\BibitemShut {NoStop}%
\bibitem [{\citenamefont {Chennakesavalu}\ \emph {et~al.}(2023)\citenamefont
  {Chennakesavalu}, \citenamefont {Toomer},\ and\ \citenamefont
  {Rotskoff}}]{chennakesavalu2023ensuring}%
  \BibitemOpen
  \bibfield  {author} {\bibinfo {author} {\bibfnamefont {S.}~\bibnamefont
  {Chennakesavalu}}, \bibinfo {author} {\bibfnamefont {D.~J.}\ \bibnamefont
  {Toomer}},\ and\ \bibinfo {author} {\bibfnamefont {G.~M.}\ \bibnamefont
  {Rotskoff}},\ }\bibfield  {title} {\bibinfo {title} {Ensuring thermodynamic
  consistency with invertible coarse-raining},\ }\href@noop {} {\bibfield
  {journal} {\bibinfo  {journal} {J. Chem. Phys.}\ }\textbf {\bibinfo {volume}
  {158}},\ \bibinfo {pages} {124126} (\bibinfo {year} {2023})}\BibitemShut
  {NoStop}%
\bibitem [{\citenamefont {Kr\"amer}\ \emph {et~al.}(2023)\citenamefont
  {Kr\"amer}, \citenamefont {Durumeric}, \citenamefont {Charron}, \citenamefont
  {Chen}, \citenamefont {Clementi},\ and\ \citenamefont
  {No{\'e}}}]{kraemer2023statistically}%
  \BibitemOpen
  \bibfield  {author} {\bibinfo {author} {\bibfnamefont {A.}~\bibnamefont
  {Kr\"amer}}, \bibinfo {author} {\bibfnamefont {A.~E.}\ \bibnamefont
  {Durumeric}}, \bibinfo {author} {\bibfnamefont {N.~E.}\ \bibnamefont
  {Charron}}, \bibinfo {author} {\bibfnamefont {Y.}~\bibnamefont {Chen}},
  \bibinfo {author} {\bibfnamefont {C.}~\bibnamefont {Clementi}},\ and\
  \bibinfo {author} {\bibfnamefont {F.}~\bibnamefont {No{\'e}}},\ }\bibfield
  {title} {\bibinfo {title} {Statistically optimal force aggregation for
  coarse-graining molecular dynamics},\ }\href@noop {} {\bibfield  {journal}
  {\bibinfo  {journal} {J. Chem. Phys. Lett.}\ }\textbf {\bibinfo {volume}
  {14}},\ \bibinfo {pages} {3970} (\bibinfo {year} {2023})}\BibitemShut
  {NoStop}%
\bibitem [{\citenamefont {Wellawatte}\ \emph {et~al.}(2023)\citenamefont
  {Wellawatte}, \citenamefont {Hocky},\ and\ \citenamefont
  {White}}]{wellawatte2023neural}%
  \BibitemOpen
  \bibfield  {author} {\bibinfo {author} {\bibfnamefont {G.~P.}\ \bibnamefont
  {Wellawatte}}, \bibinfo {author} {\bibfnamefont {G.~M.}\ \bibnamefont
  {Hocky}},\ and\ \bibinfo {author} {\bibfnamefont {A.~D.}\ \bibnamefont
  {White}},\ }\bibfield  {title} {\bibinfo {title} {Neural potentials of
  proteins extrapolate beyond training data},\ }\href@noop {} {\bibfield
  {journal} {\bibinfo  {journal} {J. Chem. Phys.}\ }\textbf {\bibinfo {volume}
  {159}} (\bibinfo {year} {2023})}\BibitemShut {NoStop}%
\bibitem [{\citenamefont {Airas}\ \emph {et~al.}(2023)\citenamefont {Airas},
  \citenamefont {Ding},\ and\ \citenamefont {Zhang}}]{airas2023transferable}%
  \BibitemOpen
  \bibfield  {author} {\bibinfo {author} {\bibfnamefont {J.}~\bibnamefont
  {Airas}}, \bibinfo {author} {\bibfnamefont {X.}~\bibnamefont {Ding}},\ and\
  \bibinfo {author} {\bibfnamefont {B.}~\bibnamefont {Zhang}},\ }\bibfield
  {title} {\bibinfo {title} {Transferable implicit solvation via contrastive
  learning of graph neural networks},\ }\href
  {http://dx.doi.org/10.1021/acscentsci.3c01160} {\bibfield  {journal}
  {\bibinfo  {journal} {ACS Cent. Sci.}\ }\textbf {\bibinfo {volume} {9}},\
  \bibinfo {pages} {2286–2297} (\bibinfo {year} {2023})}\BibitemShut
  {NoStop}%
\bibitem [{\citenamefont {Arts}\ \emph {et~al.}(2023)\citenamefont {Arts},
  \citenamefont {Garcia~Satorras}, \citenamefont {Huang}, \citenamefont
  {Z\"ugner}, \citenamefont {Federici}, \citenamefont {Clementi}, \citenamefont
  {No{\'e}}, \citenamefont {Pinsler},\ and\ \citenamefont {van~den
  Berg}}]{arts2023two}%
  \BibitemOpen
  \bibfield  {author} {\bibinfo {author} {\bibfnamefont {M.}~\bibnamefont
  {Arts}}, \bibinfo {author} {\bibfnamefont {V.}~\bibnamefont
  {Garcia~Satorras}}, \bibinfo {author} {\bibfnamefont {C.-W.}\ \bibnamefont
  {Huang}}, \bibinfo {author} {\bibfnamefont {D.}~\bibnamefont {Z\"ugner}},
  \bibinfo {author} {\bibfnamefont {M.}~\bibnamefont {Federici}}, \bibinfo
  {author} {\bibfnamefont {C.}~\bibnamefont {Clementi}}, \bibinfo {author}
  {\bibfnamefont {F.}~\bibnamefont {No{\'e}}}, \bibinfo {author} {\bibfnamefont
  {R.}~\bibnamefont {Pinsler}},\ and\ \bibinfo {author} {\bibfnamefont
  {R.}~\bibnamefont {van~den Berg}},\ }\bibfield  {title} {\bibinfo {title}
  {Two for one: Diffusion models and force fields for coarse-grained molecular
  dynamics},\ }\href@noop {} {\bibfield  {journal} {\bibinfo  {journal} {J.
  Chem. Theory Comput.}\ }\textbf {\bibinfo {volume} {19}},\ \bibinfo {pages}
  {6151} (\bibinfo {year} {2023})}\BibitemShut {NoStop}%
\bibitem [{\citenamefont {Hyv{\"a}rinen}\ and\ \citenamefont
  {Dayan}(2005)}]{hyvarinen2005estimation}%
  \BibitemOpen
  \bibfield  {author} {\bibinfo {author} {\bibfnamefont {A.}~\bibnamefont
  {Hyv{\"a}rinen}}\ and\ \bibinfo {author} {\bibfnamefont {P.}~\bibnamefont
  {Dayan}},\ }\bibfield  {title} {\bibinfo {title} {Estimation of
  non-normalized statistical models by score matching.},\ }\href@noop {}
  {\bibfield  {journal} {\bibinfo  {journal} {Journal of Machine Learning
  Research}\ }\textbf {\bibinfo {volume} {6}} (\bibinfo {year}
  {2005})}\BibitemShut {NoStop}%
\bibitem [{\citenamefont {Vincent}(2011)}]{vincent2011connection}%
  \BibitemOpen
  \bibfield  {author} {\bibinfo {author} {\bibfnamefont {P.}~\bibnamefont
  {Vincent}},\ }\bibfield  {title} {\bibinfo {title} {A connection between
  score matching and denoising autoencoders},\ }\href@noop {} {\bibfield
  {journal} {\bibinfo  {journal} {Neural Comput.}\ }\textbf {\bibinfo {volume}
  {23}},\ \bibinfo {pages} {1661} (\bibinfo {year} {2011})}\BibitemShut
  {NoStop}%
\bibitem [{\citenamefont {Song}\ and\ \citenamefont
  {Ermon}(2019)}]{DenoisingScoreMatchig}%
  \BibitemOpen
  \bibfield  {author} {\bibinfo {author} {\bibfnamefont {Y.}~\bibnamefont
  {Song}}\ and\ \bibinfo {author} {\bibfnamefont {S.}~\bibnamefont {Ermon}},\
  }\href@noop {} {\bibinfo {title} {Generative modeling by estimating gradients
  of the data distribution}} (\bibinfo {year} {2019})\BibitemShut {NoStop}%
\bibitem [{\citenamefont {Sohl-Dickstein}\ \emph {et~al.}(2015)\citenamefont
  {Sohl-Dickstein}, \citenamefont {Weiss}, \citenamefont {Maheswaranathan},\
  and\ \citenamefont {Ganguli}}]{sohl2015deep}%
  \BibitemOpen
  \bibfield  {author} {\bibinfo {author} {\bibfnamefont {J.}~\bibnamefont
  {Sohl-Dickstein}}, \bibinfo {author} {\bibfnamefont {E.}~\bibnamefont
  {Weiss}}, \bibinfo {author} {\bibfnamefont {N.}~\bibnamefont
  {Maheswaranathan}},\ and\ \bibinfo {author} {\bibfnamefont {S.}~\bibnamefont
  {Ganguli}},\ }\bibfield  {title} {\bibinfo {title} {Deep unsupervised
  learning using nonequilibrium thermodynamics},\ }in\ \href@noop {} {\emph
  {\bibinfo {booktitle} {Int. Conf. Mach. Learn.}}}\ (\bibinfo {organization}
  {PMLR},\ \bibinfo {year} {2015})\ pp.\ \bibinfo {pages}
  {2256--2265}\BibitemShut {NoStop}%
\bibitem [{\citenamefont {Ho}\ \emph {et~al.}(2020)\citenamefont {Ho},
  \citenamefont {Jain},\ and\ \citenamefont {Abbeel}}]{ho2020denoising}%
  \BibitemOpen
  \bibfield  {author} {\bibinfo {author} {\bibfnamefont {J.}~\bibnamefont
  {Ho}}, \bibinfo {author} {\bibfnamefont {A.}~\bibnamefont {Jain}},\ and\
  \bibinfo {author} {\bibfnamefont {P.}~\bibnamefont {Abbeel}},\ }\bibfield
  {title} {\bibinfo {title} {Denoising diffusion probabilistic models},\
  }\href@noop {} {\bibfield  {journal} {\bibinfo  {journal} {Adv. Neural Inf.
  Process. Syst.}\ }\textbf {\bibinfo {volume} {33}},\ \bibinfo {pages} {6840}
  (\bibinfo {year} {2020})}\BibitemShut {NoStop}%
\bibitem [{\citenamefont {Song}\ \emph {et~al.}(2020)\citenamefont {Song},
  \citenamefont {Sohl-Dickstein}, \citenamefont {Kingma}, \citenamefont
  {Kumar}, \citenamefont {Ermon},\ and\ \citenamefont {Poole}}]{song2020score}%
  \BibitemOpen
  \bibfield  {author} {\bibinfo {author} {\bibfnamefont {Y.}~\bibnamefont
  {Song}}, \bibinfo {author} {\bibfnamefont {J.}~\bibnamefont
  {Sohl-Dickstein}}, \bibinfo {author} {\bibfnamefont {D.~P.}\ \bibnamefont
  {Kingma}}, \bibinfo {author} {\bibfnamefont {A.}~\bibnamefont {Kumar}},
  \bibinfo {author} {\bibfnamefont {S.}~\bibnamefont {Ermon}},\ and\ \bibinfo
  {author} {\bibfnamefont {B.}~\bibnamefont {Poole}},\ }\bibfield  {title}
  {\bibinfo {title} {Score-based generative modeling through stochastic
  differential equations},\ }\href@noop {} {\bibfield  {journal} {\bibinfo
  {journal} {arXiv preprint arXiv:2011.13456}\ } (\bibinfo {year}
  {2020})}\BibitemShut {NoStop}%
\bibitem [{\citenamefont {Barua}\ \emph {et~al.}(2008)\citenamefont {Barua},
  \citenamefont {Lin}, \citenamefont {Williams}, \citenamefont {Kummler},
  \citenamefont {Neidigh},\ and\ \citenamefont {Andersen}}]{barua2008trp}%
  \BibitemOpen
  \bibfield  {author} {\bibinfo {author} {\bibfnamefont {B.}~\bibnamefont
  {Barua}}, \bibinfo {author} {\bibfnamefont {J.~C.}\ \bibnamefont {Lin}},
  \bibinfo {author} {\bibfnamefont {V.~D.}\ \bibnamefont {Williams}}, \bibinfo
  {author} {\bibfnamefont {P.}~\bibnamefont {Kummler}}, \bibinfo {author}
  {\bibfnamefont {J.~W.}\ \bibnamefont {Neidigh}},\ and\ \bibinfo {author}
  {\bibfnamefont {N.~H.}\ \bibnamefont {Andersen}},\ }\bibfield  {title}
  {\bibinfo {title} {The trp-cage: optimizing the stability of a globular
  miniprotein},\ }\href@noop {} {\bibfield  {journal} {\bibinfo  {journal}
  {Protein Eng. Des. Sel.}\ }\textbf {\bibinfo {volume} {21}},\ \bibinfo
  {pages} {171} (\bibinfo {year} {2008})}\BibitemShut {NoStop}%
\bibitem [{\citenamefont {Cho}\ \emph {et~al.}(2004)\citenamefont {Cho},
  \citenamefont {Sato},\ and\ \citenamefont {Raleigh}}]{cho2004ntl9}%
  \BibitemOpen
  \bibfield  {author} {\bibinfo {author} {\bibfnamefont {J.-H.}\ \bibnamefont
  {Cho}}, \bibinfo {author} {\bibfnamefont {S.}~\bibnamefont {Sato}},\ and\
  \bibinfo {author} {\bibfnamefont {D.~P.}\ \bibnamefont {Raleigh}},\
  }\bibfield  {title} {\bibinfo {title} {Thermodynamics and kinetics of
  non-native interactions in protein folding: A single point mutant
  significantly stabilizes the n-terminal domain of l9 by modulating non-native
  interactions in the denatured state},\ }\href@noop {} {\bibfield  {journal}
  {\bibinfo  {journal} {J. Mol. Biol.}\ }\textbf {\bibinfo {volume} {338}},\
  \bibinfo {pages} {827} (\bibinfo {year} {2004})}\BibitemShut {NoStop}%
\bibitem [{\citenamefont {Ciccotti}\ \emph {et~al.}(2005)\citenamefont
  {Ciccotti}, \citenamefont {Kapral},\ and\ \citenamefont
  {Vanden-Eijnden}}]{Ciccotti2005}%
  \BibitemOpen
  \bibfield  {author} {\bibinfo {author} {\bibfnamefont {G.}~\bibnamefont
  {Ciccotti}}, \bibinfo {author} {\bibfnamefont {R.}~\bibnamefont {Kapral}},\
  and\ \bibinfo {author} {\bibfnamefont {E.}~\bibnamefont {Vanden-Eijnden}},\
  }\bibfield  {title} {\bibinfo {title} {Blue moon sampling, vectorial reaction
  coordinates, and unbiased constrained dynamics},\ }\href
  {https://doi.org/10.1002/cphc.200400669} {\bibfield  {journal} {\bibinfo
  {journal} {ChemPhysChem}\ }\textbf {\bibinfo {volume} {6}},\ \bibinfo {pages}
  {1809} (\bibinfo {year} {2005})}\BibitemShut {NoStop}%
\bibitem [{\citenamefont {Thaler}\ and\ \citenamefont
  {Zavadlav}(2021)}]{thaler2021learning}%
  \BibitemOpen
  \bibfield  {author} {\bibinfo {author} {\bibfnamefont {S.}~\bibnamefont
  {Thaler}}\ and\ \bibinfo {author} {\bibfnamefont {J.}~\bibnamefont
  {Zavadlav}},\ }\bibfield  {title} {\bibinfo {title} {Learning neural network
  potentials from experimental data via differentiable trajectory
  reweighting},\ }\href@noop {} {\bibfield  {journal} {\bibinfo  {journal}
  {Nat. Commun.}\ }\textbf {\bibinfo {volume} {12}},\ \bibinfo {pages} {6884}
  (\bibinfo {year} {2021})}\BibitemShut {NoStop}%
\bibitem [{\citenamefont {Thaler}\ \emph {et~al.}(2022)\citenamefont {Thaler},
  \citenamefont {Stupp},\ and\ \citenamefont {Zavadlav}}]{thaler2022deep}%
  \BibitemOpen
  \bibfield  {author} {\bibinfo {author} {\bibfnamefont {S.}~\bibnamefont
  {Thaler}}, \bibinfo {author} {\bibfnamefont {M.}~\bibnamefont {Stupp}},\ and\
  \bibinfo {author} {\bibfnamefont {J.}~\bibnamefont {Zavadlav}},\ }\bibfield
  {title} {\bibinfo {title} {Deep coarse-grained potentials via relative
  entropy minimization},\ }\href@noop {} {\bibfield  {journal} {\bibinfo
  {journal} {J. Chem. Phys.}\ }\textbf {\bibinfo {volume} {157}},\ \bibinfo
  {pages} {244103} (\bibinfo {year} {2022})}\BibitemShut {NoStop}%
\bibitem [{\citenamefont {Yang}\ \emph {et~al.}(2023)\citenamefont {Yang},
  \citenamefont {Zhang}, \citenamefont {Song}, \citenamefont {Hong},
  \citenamefont {Xu}, \citenamefont {Zhao}, \citenamefont {Zhang},
  \citenamefont {Cui},\ and\ \citenamefont {Yang}}]{yang2023diffusion}%
  \BibitemOpen
  \bibfield  {author} {\bibinfo {author} {\bibfnamefont {L.}~\bibnamefont
  {Yang}}, \bibinfo {author} {\bibfnamefont {Z.}~\bibnamefont {Zhang}},
  \bibinfo {author} {\bibfnamefont {Y.}~\bibnamefont {Song}}, \bibinfo {author}
  {\bibfnamefont {S.}~\bibnamefont {Hong}}, \bibinfo {author} {\bibfnamefont
  {R.}~\bibnamefont {Xu}}, \bibinfo {author} {\bibfnamefont {Y.}~\bibnamefont
  {Zhao}}, \bibinfo {author} {\bibfnamefont {W.}~\bibnamefont {Zhang}},
  \bibinfo {author} {\bibfnamefont {B.}~\bibnamefont {Cui}},\ and\ \bibinfo
  {author} {\bibfnamefont {M.-H.}\ \bibnamefont {Yang}},\ }\bibfield  {title}
  {\bibinfo {title} {Diffusion models: A comprehensive survey of methods and
  applications},\ }\href@noop {} {\bibfield  {journal} {\bibinfo  {journal}
  {ACM Computing Surveys}\ }\textbf {\bibinfo {volume} {56}},\ \bibinfo {pages}
  {1} (\bibinfo {year} {2023})}\BibitemShut {NoStop}%
\bibitem [{\citenamefont {Zheng}\ \emph {et~al.}(2024)\citenamefont {Zheng},
  \citenamefont {He}, \citenamefont {Liu}, \citenamefont {Shi}, \citenamefont
  {Lu}, \citenamefont {Feng}, \citenamefont {Ju}, \citenamefont {Wang},
  \citenamefont {Zhu}, \citenamefont {Min}, \citenamefont {Zhang},
  \citenamefont {Tang}, \citenamefont {Hao}, \citenamefont {Jin}, \citenamefont
  {Chen}, \citenamefont {No{\'e}}, \citenamefont {Liu},\ and\ \citenamefont
  {Liu}}]{Zheng2024}%
  \BibitemOpen
  \bibfield  {author} {\bibinfo {author} {\bibfnamefont {S.}~\bibnamefont
  {Zheng}}, \bibinfo {author} {\bibfnamefont {J.}~\bibnamefont {He}}, \bibinfo
  {author} {\bibfnamefont {C.}~\bibnamefont {Liu}}, \bibinfo {author}
  {\bibfnamefont {Y.}~\bibnamefont {Shi}}, \bibinfo {author} {\bibfnamefont
  {Z.}~\bibnamefont {Lu}}, \bibinfo {author} {\bibfnamefont {W.}~\bibnamefont
  {Feng}}, \bibinfo {author} {\bibfnamefont {F.}~\bibnamefont {Ju}}, \bibinfo
  {author} {\bibfnamefont {J.}~\bibnamefont {Wang}}, \bibinfo {author}
  {\bibfnamefont {J.}~\bibnamefont {Zhu}}, \bibinfo {author} {\bibfnamefont
  {Y.}~\bibnamefont {Min}}, \bibinfo {author} {\bibfnamefont {H.}~\bibnamefont
  {Zhang}}, \bibinfo {author} {\bibfnamefont {S.}~\bibnamefont {Tang}},
  \bibinfo {author} {\bibfnamefont {H.}~\bibnamefont {Hao}}, \bibinfo {author}
  {\bibfnamefont {P.}~\bibnamefont {Jin}}, \bibinfo {author} {\bibfnamefont
  {C.}~\bibnamefont {Chen}}, \bibinfo {author} {\bibfnamefont {F.}~\bibnamefont
  {No{\'e}}}, \bibinfo {author} {\bibfnamefont {H.}~\bibnamefont {Liu}},\ and\
  \bibinfo {author} {\bibfnamefont {T.-Y.}\ \bibnamefont {Liu}},\ }\bibfield
  {title} {\bibinfo {title} {Predicting equilibrium distributions for molecular
  systems with deep learning},\ }\href
  {https://doi.org/10.1038/s42256-024-00837-3} {\bibfield  {journal} {\bibinfo
  {journal} {Nat. Mach. Intell.}\ }\textbf {\bibinfo {volume} {6}},\ \bibinfo
  {pages} {558–567} (\bibinfo {year} {2024})}\BibitemShut {NoStop}%
\bibitem [{\citenamefont {De~Bortoli}\ \emph {et~al.}(2024)\citenamefont
  {De~Bortoli}, \citenamefont {Hutchinson}, \citenamefont {Wirnsberger},\ and\
  \citenamefont {Doucet}}]{de2024target}%
  \BibitemOpen
  \bibfield  {author} {\bibinfo {author} {\bibfnamefont {V.}~\bibnamefont
  {De~Bortoli}}, \bibinfo {author} {\bibfnamefont {M.}~\bibnamefont
  {Hutchinson}}, \bibinfo {author} {\bibfnamefont {P.}~\bibnamefont
  {Wirnsberger}},\ and\ \bibinfo {author} {\bibfnamefont {A.}~\bibnamefont
  {Doucet}},\ }\bibfield  {title} {\bibinfo {title} {Target score matching},\
  }\href@noop {} {\bibfield  {journal} {\bibinfo  {journal} {arXiv preprint
  arXiv:2402.08667}\ } (\bibinfo {year} {2024})}\BibitemShut {NoStop}%
\bibitem [{\citenamefont {Krishna}\ \emph {et~al.}(2024)\citenamefont
  {Krishna}, \citenamefont {Wang}, \citenamefont {Ahern}, \citenamefont
  {Sturmfels}, \citenamefont {Venkatesh}, \citenamefont {Kalvet}, \citenamefont
  {Lee}, \citenamefont {Morey-Burrows}, \citenamefont {Anishchenko},
  \citenamefont {Humphreys} \emph {et~al.}}]{krishna2024generalized}%
  \BibitemOpen
  \bibfield  {author} {\bibinfo {author} {\bibfnamefont {R.}~\bibnamefont
  {Krishna}}, \bibinfo {author} {\bibfnamefont {J.}~\bibnamefont {Wang}},
  \bibinfo {author} {\bibfnamefont {W.}~\bibnamefont {Ahern}}, \bibinfo
  {author} {\bibfnamefont {P.}~\bibnamefont {Sturmfels}}, \bibinfo {author}
  {\bibfnamefont {P.}~\bibnamefont {Venkatesh}}, \bibinfo {author}
  {\bibfnamefont {I.}~\bibnamefont {Kalvet}}, \bibinfo {author} {\bibfnamefont
  {G.~R.}\ \bibnamefont {Lee}}, \bibinfo {author} {\bibfnamefont {F.~S.}\
  \bibnamefont {Morey-Burrows}}, \bibinfo {author} {\bibfnamefont
  {I.}~\bibnamefont {Anishchenko}}, \bibinfo {author} {\bibfnamefont {I.~R.}\
  \bibnamefont {Humphreys}}, \emph {et~al.},\ }\bibfield  {title} {\bibinfo
  {title} {Generalized biomolecular modeling and design with rosettafold
  all-atom},\ }\href@noop {} {\bibfield  {journal} {\bibinfo  {journal}
  {Science}\ }\textbf {\bibinfo {volume} {384}},\ \bibinfo {pages} {eadl2528}
  (\bibinfo {year} {2024})}\BibitemShut {NoStop}%
\bibitem [{\citenamefont {Prinz}\ \emph {et~al.}(2011)\citenamefont {Prinz},
  \citenamefont {Wu}, \citenamefont {Sarich}, \citenamefont {Keller},
  \citenamefont {Senne}, \citenamefont {Held}, \citenamefont {Chodera},
  \citenamefont {Sch{\"u}tte},\ and\ \citenamefont
  {No{\'e}}}]{prinz2011markov}%
  \BibitemOpen
  \bibfield  {author} {\bibinfo {author} {\bibfnamefont {J.-H.}\ \bibnamefont
  {Prinz}}, \bibinfo {author} {\bibfnamefont {H.}~\bibnamefont {Wu}}, \bibinfo
  {author} {\bibfnamefont {M.}~\bibnamefont {Sarich}}, \bibinfo {author}
  {\bibfnamefont {B.}~\bibnamefont {Keller}}, \bibinfo {author} {\bibfnamefont
  {M.}~\bibnamefont {Senne}}, \bibinfo {author} {\bibfnamefont
  {M.}~\bibnamefont {Held}}, \bibinfo {author} {\bibfnamefont {J.~D.}\
  \bibnamefont {Chodera}}, \bibinfo {author} {\bibfnamefont {C.}~\bibnamefont
  {Sch{\"u}tte}},\ and\ \bibinfo {author} {\bibfnamefont {F.}~\bibnamefont
  {No{\'e}}},\ }\bibfield  {title} {\bibinfo {title} {Markov models of
  molecular kinetics: Generation and validation},\ }\href@noop {} {\bibfield
  {journal} {\bibinfo  {journal} {J. Chem. Phys.}\ }\textbf {\bibinfo {volume}
  {134}} (\bibinfo {year} {2011})}\BibitemShut {NoStop}%
\bibitem [{\citenamefont {P{\'e}rez-Hern{\'a}ndez}\ \emph
  {et~al.}(2013)\citenamefont {P{\'e}rez-Hern{\'a}ndez}, \citenamefont {Paul},
  \citenamefont {Giorgino}, \citenamefont {De~Fabritiis},\ and\ \citenamefont
  {No{\'e}}}]{Perez_JChemPhys2013}%
  \BibitemOpen
  \bibfield  {author} {\bibinfo {author} {\bibfnamefont {G.}~\bibnamefont
  {P{\'e}rez-Hern{\'a}ndez}}, \bibinfo {author} {\bibfnamefont
  {F.}~\bibnamefont {Paul}}, \bibinfo {author} {\bibfnamefont {T.}~\bibnamefont
  {Giorgino}}, \bibinfo {author} {\bibfnamefont {G.}~\bibnamefont
  {De~Fabritiis}},\ and\ \bibinfo {author} {\bibfnamefont {F.}~\bibnamefont
  {No{\'e}}},\ }\bibfield  {title} {\bibinfo {title} {Identification of slow
  molecular order parameters for markov model construction},\ }\href@noop {}
  {\bibfield  {journal} {\bibinfo  {journal} {J. Chem. Phys.}\ }\textbf
  {\bibinfo {volume} {139}},\ \bibinfo {pages} {015102} (\bibinfo {year}
  {2013})}\BibitemShut {NoStop}%
\bibitem [{\citenamefont {Schwantes}\ and\ \citenamefont
  {Pande}(2013)}]{Schwantes_JChemTheoryComput2013}%
  \BibitemOpen
  \bibfield  {author} {\bibinfo {author} {\bibfnamefont {C.~R.}\ \bibnamefont
  {Schwantes}}\ and\ \bibinfo {author} {\bibfnamefont {V.~S.}\ \bibnamefont
  {Pande}},\ }\bibfield  {title} {\bibinfo {title} {Improvements in markov
  state model construction reveal many non-native interactions in the folding
  of ntl9},\ }\href@noop {} {\bibfield  {journal} {\bibinfo  {journal} {J.
  Chem. Theory Comput.}\ }\textbf {\bibinfo {volume} {9}},\ \bibinfo {pages}
  {2000} (\bibinfo {year} {2013})}\BibitemShut {NoStop}%
\bibitem [{\citenamefont {Schwantes}\ \emph {et~al.}(2016)\citenamefont
  {Schwantes}, \citenamefont {Shukla},\ and\ \citenamefont
  {Pande}}]{Schwantes_BiophysJ2016}%
  \BibitemOpen
  \bibfield  {author} {\bibinfo {author} {\bibfnamefont {C.~R.}\ \bibnamefont
  {Schwantes}}, \bibinfo {author} {\bibfnamefont {D.}~\bibnamefont {Shukla}},\
  and\ \bibinfo {author} {\bibfnamefont {V.~S.}\ \bibnamefont {Pande}},\
  }\bibfield  {title} {\bibinfo {title} {Markov state models and tica reveal a
  nonnative folding nucleus in simulations of nug2},\ }\href
  {https://doi.org/https://doi.org/10.1016/j.bpj.2016.03.026} {\bibfield
  {journal} {\bibinfo  {journal} {Biophys. J.}\ }\textbf {\bibinfo {volume}
  {110}},\ \bibinfo {pages} {1716} (\bibinfo {year} {2016})}\BibitemShut
  {NoStop}%
\bibitem [{\citenamefont {Unke}\ \emph {et~al.}(2021)\citenamefont {Unke},
  \citenamefont {Chmiela}, \citenamefont {Sauceda}, \citenamefont {Gastegger},
  \citenamefont {Poltavsky}, \citenamefont {Sch\"utt}, \citenamefont
  {Tkatchenko},\ and\ \citenamefont {M\"uller}}]{unke2021machine}%
  \BibitemOpen
  \bibfield  {author} {\bibinfo {author} {\bibfnamefont {O.~T.}\ \bibnamefont
  {Unke}}, \bibinfo {author} {\bibfnamefont {S.}~\bibnamefont {Chmiela}},
  \bibinfo {author} {\bibfnamefont {H.~E.}\ \bibnamefont {Sauceda}}, \bibinfo
  {author} {\bibfnamefont {M.}~\bibnamefont {Gastegger}}, \bibinfo {author}
  {\bibfnamefont {I.}~\bibnamefont {Poltavsky}}, \bibinfo {author}
  {\bibfnamefont {K.~T.}\ \bibnamefont {Sch\"utt}}, \bibinfo {author}
  {\bibfnamefont {A.}~\bibnamefont {Tkatchenko}},\ and\ \bibinfo {author}
  {\bibfnamefont {K.-R.}\ \bibnamefont {M\"uller}},\ }\bibfield  {title}
  {\bibinfo {title} {Machine learning force fields},\ }\href@noop {} {\bibfield
   {journal} {\bibinfo  {journal} {Chem. Rev.}\ }\textbf {\bibinfo {volume}
  {121}},\ \bibinfo {pages} {10142} (\bibinfo {year} {2021})}\BibitemShut
  {NoStop}%
\bibitem [{\citenamefont {Duschatko}\ \emph
  {et~al.}(2024{\natexlab{a}})\citenamefont {Duschatko}, \citenamefont {Fu},
  \citenamefont {Owen}, \citenamefont {Xie}, \citenamefont {Musaelian},
  \citenamefont {Jaakkola},\ and\ \citenamefont
  {Kozinsky}}]{duschatko2024thermodynamically}%
  \BibitemOpen
  \bibfield  {author} {\bibinfo {author} {\bibfnamefont {B.~R.}\ \bibnamefont
  {Duschatko}}, \bibinfo {author} {\bibfnamefont {X.}~\bibnamefont {Fu}},
  \bibinfo {author} {\bibfnamefont {C.}~\bibnamefont {Owen}}, \bibinfo {author}
  {\bibfnamefont {Y.}~\bibnamefont {Xie}}, \bibinfo {author} {\bibfnamefont
  {A.}~\bibnamefont {Musaelian}}, \bibinfo {author} {\bibfnamefont
  {T.}~\bibnamefont {Jaakkola}},\ and\ \bibinfo {author} {\bibfnamefont
  {B.}~\bibnamefont {Kozinsky}},\ }\bibfield  {title} {\bibinfo {title}
  {Thermodynamically informed multimodal learning of high-dimensional free
  energy models in molecular coarse graining},\ }\href@noop {} {\bibfield
  {journal} {\bibinfo  {journal} {arXiv preprint arXiv:2405.19386}\ } (\bibinfo
  {year} {2024}{\natexlab{a}})}\BibitemShut {NoStop}%
\bibitem [{\citenamefont {Sch{\"u}tt}\ \emph {et~al.}(2018)\citenamefont
  {Sch{\"u}tt}, \citenamefont {Sauceda}, \citenamefont {Kindermans},
  \citenamefont {Tkatchenko},\ and\ \citenamefont
  {M{\"u}ller}}]{schutt2018schnet}%
  \BibitemOpen
  \bibfield  {author} {\bibinfo {author} {\bibfnamefont {K.~T.}\ \bibnamefont
  {Sch{\"u}tt}}, \bibinfo {author} {\bibfnamefont {H.~E.}\ \bibnamefont
  {Sauceda}}, \bibinfo {author} {\bibfnamefont {P.-J.}\ \bibnamefont
  {Kindermans}}, \bibinfo {author} {\bibfnamefont {A.}~\bibnamefont
  {Tkatchenko}},\ and\ \bibinfo {author} {\bibfnamefont {K.-R.}\ \bibnamefont
  {M{\"u}ller}},\ }\bibfield  {title} {\bibinfo {title} {Schnet--a deep
  learning architecture for molecules and materials},\ }\href@noop {}
  {\bibfield  {journal} {\bibinfo  {journal} {J. Chem. Phys.}\ }\textbf
  {\bibinfo {volume} {148}},\ \bibinfo {pages} {241722} (\bibinfo {year}
  {2018})}\BibitemShut {NoStop}%
\bibitem [{\citenamefont {Harris}\ \emph {et~al.}(2020)\citenamefont {Harris},
  \citenamefont {Millman}, \citenamefont {van~der Walt}, \citenamefont
  {Gommers}, \citenamefont {Virtanen}, \citenamefont {Cournapeau},
  \citenamefont {Wieser}, \citenamefont {Taylor}, \citenamefont {Berg},
  \citenamefont {Smith}, \citenamefont {Kern}, \citenamefont {Picus},
  \citenamefont {Hoyer}, \citenamefont {van Kerkwijk}, \citenamefont {Brett},
  \citenamefont {Haldane}, \citenamefont {del R{\'{i}}o}, \citenamefont
  {Wiebe}, \citenamefont {Peterson}, \citenamefont {G{\'{e}}rard-Marchant},
  \citenamefont {Sheppard}, \citenamefont {Reddy}, \citenamefont {Weckesser},
  \citenamefont {Abbasi}, \citenamefont {Gohlke},\ and\ \citenamefont
  {Oliphant}}]{harris2020array}%
  \BibitemOpen
  \bibfield  {author} {\bibinfo {author} {\bibfnamefont {C.~R.}\ \bibnamefont
  {Harris}}, \bibinfo {author} {\bibfnamefont {K.~J.}\ \bibnamefont {Millman}},
  \bibinfo {author} {\bibfnamefont {S.~J.}\ \bibnamefont {van~der Walt}},
  \bibinfo {author} {\bibfnamefont {R.}~\bibnamefont {Gommers}}, \bibinfo
  {author} {\bibfnamefont {P.}~\bibnamefont {Virtanen}}, \bibinfo {author}
  {\bibfnamefont {D.}~\bibnamefont {Cournapeau}}, \bibinfo {author}
  {\bibfnamefont {E.}~\bibnamefont {Wieser}}, \bibinfo {author} {\bibfnamefont
  {J.}~\bibnamefont {Taylor}}, \bibinfo {author} {\bibfnamefont
  {S.}~\bibnamefont {Berg}}, \bibinfo {author} {\bibfnamefont {N.~J.}\
  \bibnamefont {Smith}}, \bibinfo {author} {\bibfnamefont {R.}~\bibnamefont
  {Kern}}, \bibinfo {author} {\bibfnamefont {M.}~\bibnamefont {Picus}},
  \bibinfo {author} {\bibfnamefont {S.}~\bibnamefont {Hoyer}}, \bibinfo
  {author} {\bibfnamefont {M.~H.}\ \bibnamefont {van Kerkwijk}}, \bibinfo
  {author} {\bibfnamefont {M.}~\bibnamefont {Brett}}, \bibinfo {author}
  {\bibfnamefont {A.}~\bibnamefont {Haldane}}, \bibinfo {author} {\bibfnamefont
  {J.~F.}\ \bibnamefont {del R{\'{i}}o}}, \bibinfo {author} {\bibfnamefont
  {M.}~\bibnamefont {Wiebe}}, \bibinfo {author} {\bibfnamefont
  {P.}~\bibnamefont {Peterson}}, \bibinfo {author} {\bibfnamefont
  {P.}~\bibnamefont {G{\'{e}}rard-Marchant}}, \bibinfo {author} {\bibfnamefont
  {K.}~\bibnamefont {Sheppard}}, \bibinfo {author} {\bibfnamefont
  {T.}~\bibnamefont {Reddy}}, \bibinfo {author} {\bibfnamefont
  {W.}~\bibnamefont {Weckesser}}, \bibinfo {author} {\bibfnamefont
  {H.}~\bibnamefont {Abbasi}}, \bibinfo {author} {\bibfnamefont
  {C.}~\bibnamefont {Gohlke}},\ and\ \bibinfo {author} {\bibfnamefont {T.~E.}\
  \bibnamefont {Oliphant}},\ }\bibfield  {title} {\bibinfo {title} {Array
  programming with {NumPy}},\ }\href
  {https://doi.org/10.1038/s41586-020-2649-2} {\bibfield  {journal} {\bibinfo
  {journal} {Nature}\ }\textbf {\bibinfo {volume} {585}},\ \bibinfo {pages}
  {357} (\bibinfo {year} {2020})}\BibitemShut {NoStop}%
\bibitem [{\citenamefont {Bradbury}\ \emph {et~al.}(2018)\citenamefont
  {Bradbury}, \citenamefont {Frostig}, \citenamefont {Hawkins}, \citenamefont
  {Johnson}, \citenamefont {Leary}, \citenamefont {Maclaurin}, \citenamefont
  {Necula}, \citenamefont {Paszke}, \citenamefont {Vander{P}las}, \citenamefont
  {Wanderman-{M}ilne},\ and\ \citenamefont {Zhang}}]{jax2018github}%
  \BibitemOpen
  \bibfield  {author} {\bibinfo {author} {\bibfnamefont {J.}~\bibnamefont
  {Bradbury}}, \bibinfo {author} {\bibfnamefont {R.}~\bibnamefont {Frostig}},
  \bibinfo {author} {\bibfnamefont {P.}~\bibnamefont {Hawkins}}, \bibinfo
  {author} {\bibfnamefont {M.~J.}\ \bibnamefont {Johnson}}, \bibinfo {author}
  {\bibfnamefont {C.}~\bibnamefont {Leary}}, \bibinfo {author} {\bibfnamefont
  {D.}~\bibnamefont {Maclaurin}}, \bibinfo {author} {\bibfnamefont
  {G.}~\bibnamefont {Necula}}, \bibinfo {author} {\bibfnamefont
  {A.}~\bibnamefont {Paszke}}, \bibinfo {author} {\bibfnamefont
  {J.}~\bibnamefont {Vander{P}las}}, \bibinfo {author} {\bibfnamefont
  {S.}~\bibnamefont {Wanderman-{M}ilne}},\ and\ \bibinfo {author}
  {\bibfnamefont {Q.}~\bibnamefont {Zhang}},\ }\href
  {http://github.com/google/jax} {\bibinfo {title} {{JAX}: composable
  transformations of {P}ython+{N}um{P}y programs}} (\bibinfo {year}
  {2018})\BibitemShut {NoStop}%
\bibitem [{\citenamefont {pandas~development team}(2020)}]{reback2020pandas}%
  \BibitemOpen
  \bibfield  {author} {\bibinfo {author} {\bibfnamefont {T.}~\bibnamefont
  {pandas~development team}},\ }\href {https://doi.org/10.5281/zenodo.3509134}
  {\bibinfo {title} {pandas-dev/pandas: Pandas}} (\bibinfo {year}
  {2020})\BibitemShut {NoStop}%
\bibitem [{\citenamefont {Virtanen}\ \emph {et~al.}(2020)\citenamefont
  {Virtanen}, \citenamefont {Gommers}, \citenamefont {Oliphant}, \citenamefont
  {Haberland}, \citenamefont {Reddy}, \citenamefont {Cournapeau}, \citenamefont
  {Burovski}, \citenamefont {Peterson}, \citenamefont {Weckesser},
  \citenamefont {Bright}, \citenamefont {{van der Walt}}, \citenamefont
  {Brett}, \citenamefont {Wilson}, \citenamefont {Millman}, \citenamefont
  {Mayorov}, \citenamefont {Nelson}, \citenamefont {Jones}, \citenamefont
  {Kern}, \citenamefont {Larson}, \citenamefont {Carey}, \citenamefont {Polat},
  \citenamefont {Feng}, \citenamefont {Moore}, \citenamefont {{VanderPlas}},
  \citenamefont {Laxalde}, \citenamefont {Perktold}, \citenamefont {Cimrman},
  \citenamefont {Henriksen}, \citenamefont {Quintero}, \citenamefont {Harris},
  \citenamefont {Archibald}, \citenamefont {Ribeiro}, \citenamefont
  {Pedregosa}, \citenamefont {{van Mulbregt}},\ and\ \citenamefont {{SciPy 1.0
  Contributors}}}]{2020SciPy-NMeth}%
  \BibitemOpen
  \bibfield  {author} {\bibinfo {author} {\bibfnamefont {P.}~\bibnamefont
  {Virtanen}}, \bibinfo {author} {\bibfnamefont {R.}~\bibnamefont {Gommers}},
  \bibinfo {author} {\bibfnamefont {T.~E.}\ \bibnamefont {Oliphant}}, \bibinfo
  {author} {\bibfnamefont {M.}~\bibnamefont {Haberland}}, \bibinfo {author}
  {\bibfnamefont {T.}~\bibnamefont {Reddy}}, \bibinfo {author} {\bibfnamefont
  {D.}~\bibnamefont {Cournapeau}}, \bibinfo {author} {\bibfnamefont
  {E.}~\bibnamefont {Burovski}}, \bibinfo {author} {\bibfnamefont
  {P.}~\bibnamefont {Peterson}}, \bibinfo {author} {\bibfnamefont
  {W.}~\bibnamefont {Weckesser}}, \bibinfo {author} {\bibfnamefont
  {J.}~\bibnamefont {Bright}}, \bibinfo {author} {\bibfnamefont {S.~J.}\
  \bibnamefont {{van der Walt}}}, \bibinfo {author} {\bibfnamefont
  {M.}~\bibnamefont {Brett}}, \bibinfo {author} {\bibfnamefont
  {J.}~\bibnamefont {Wilson}}, \bibinfo {author} {\bibfnamefont {K.~J.}\
  \bibnamefont {Millman}}, \bibinfo {author} {\bibfnamefont {N.}~\bibnamefont
  {Mayorov}}, \bibinfo {author} {\bibfnamefont {A.~R.~J.}\ \bibnamefont
  {Nelson}}, \bibinfo {author} {\bibfnamefont {E.}~\bibnamefont {Jones}},
  \bibinfo {author} {\bibfnamefont {R.}~\bibnamefont {Kern}}, \bibinfo {author}
  {\bibfnamefont {E.}~\bibnamefont {Larson}}, \bibinfo {author} {\bibfnamefont
  {C.~J.}\ \bibnamefont {Carey}}, \bibinfo {author} {\bibfnamefont
  {{\.I}.}~\bibnamefont {Polat}}, \bibinfo {author} {\bibfnamefont
  {Y.}~\bibnamefont {Feng}}, \bibinfo {author} {\bibfnamefont {E.~W.}\
  \bibnamefont {Moore}}, \bibinfo {author} {\bibfnamefont {J.}~\bibnamefont
  {{VanderPlas}}}, \bibinfo {author} {\bibfnamefont {D.}~\bibnamefont
  {Laxalde}}, \bibinfo {author} {\bibfnamefont {J.}~\bibnamefont {Perktold}},
  \bibinfo {author} {\bibfnamefont {R.}~\bibnamefont {Cimrman}}, \bibinfo
  {author} {\bibfnamefont {I.}~\bibnamefont {Henriksen}}, \bibinfo {author}
  {\bibfnamefont {E.~A.}\ \bibnamefont {Quintero}}, \bibinfo {author}
  {\bibfnamefont {C.~R.}\ \bibnamefont {Harris}}, \bibinfo {author}
  {\bibfnamefont {A.~M.}\ \bibnamefont {Archibald}}, \bibinfo {author}
  {\bibfnamefont {A.~H.}\ \bibnamefont {Ribeiro}}, \bibinfo {author}
  {\bibfnamefont {F.}~\bibnamefont {Pedregosa}}, \bibinfo {author}
  {\bibfnamefont {P.}~\bibnamefont {{van Mulbregt}}},\ and\ \bibinfo {author}
  {\bibnamefont {{SciPy 1.0 Contributors}}},\ }\bibfield  {title} {\bibinfo
  {title} {{{SciPy} 1.0: Fundamental Algorithms for Scientific Computing in
  Python}},\ }\href {https://doi.org/10.1038/s41592-019-0686-2} {\bibfield
  {journal} {\bibinfo  {journal} {Nat. Methods}\ }\textbf {\bibinfo {volume}
  {17}},\ \bibinfo {pages} {261} (\bibinfo {year} {2020})}\BibitemShut
  {NoStop}%
\bibitem [{\citenamefont {Stellato}\ \emph {et~al.}(2020)\citenamefont
  {Stellato}, \citenamefont {Banjac}, \citenamefont {Goulart}, \citenamefont
  {Bemporad},\ and\ \citenamefont {Boyd}}]{osqp}%
  \BibitemOpen
  \bibfield  {author} {\bibinfo {author} {\bibfnamefont {B.}~\bibnamefont
  {Stellato}}, \bibinfo {author} {\bibfnamefont {G.}~\bibnamefont {Banjac}},
  \bibinfo {author} {\bibfnamefont {P.}~\bibnamefont {Goulart}}, \bibinfo
  {author} {\bibfnamefont {A.}~\bibnamefont {Bemporad}},\ and\ \bibinfo
  {author} {\bibfnamefont {S.}~\bibnamefont {Boyd}},\ }\bibfield  {title}
  {\bibinfo {title} {{OSQP}: an operator splitting solver for quadratic
  programs},\ }\href {https://doi.org/10.1007/s12532-020-00179-2} {\bibfield
  {journal} {\bibinfo  {journal} {Math. Program. Comput.}\ }\textbf {\bibinfo
  {volume} {12}},\ \bibinfo {pages} {637} (\bibinfo {year} {2020})}\BibitemShut
  {NoStop}%
\bibitem [{\citenamefont {Banjac}\ \emph {et~al.}(2019)\citenamefont {Banjac},
  \citenamefont {Goulart}, \citenamefont {Stellato},\ and\ \citenamefont
  {Boyd}}]{osqp-infeasibility}%
  \BibitemOpen
  \bibfield  {author} {\bibinfo {author} {\bibfnamefont {G.}~\bibnamefont
  {Banjac}}, \bibinfo {author} {\bibfnamefont {P.}~\bibnamefont {Goulart}},
  \bibinfo {author} {\bibfnamefont {B.}~\bibnamefont {Stellato}},\ and\
  \bibinfo {author} {\bibfnamefont {S.}~\bibnamefont {Boyd}},\ }\bibfield
  {title} {\bibinfo {title} {Infeasibility detection in the alternating
  direction method of multipliers for convex optimization},\ }\href
  {https://doi.org/10.1007/s10957-019-01575-y} {\bibfield  {journal} {\bibinfo
  {journal} {J. Optim. Theory. Appl.}\ }\textbf {\bibinfo {volume} {183}},\
  \bibinfo {pages} {490} (\bibinfo {year} {2019})}\BibitemShut {NoStop}%
\bibitem [{\citenamefont {Buch}\ \emph {et~al.}(2010)\citenamefont {Buch},
  \citenamefont {Harvey}, \citenamefont {Giorgino}, \citenamefont {Anderson},\
  and\ \citenamefont {De~Fabritiis}}]{buch2010high}%
  \BibitemOpen
  \bibfield  {author} {\bibinfo {author} {\bibfnamefont {I.}~\bibnamefont
  {Buch}}, \bibinfo {author} {\bibfnamefont {M.~J.}\ \bibnamefont {Harvey}},
  \bibinfo {author} {\bibfnamefont {T.}~\bibnamefont {Giorgino}}, \bibinfo
  {author} {\bibfnamefont {D.~P.}\ \bibnamefont {Anderson}},\ and\ \bibinfo
  {author} {\bibfnamefont {G.}~\bibnamefont {De~Fabritiis}},\ }\bibfield
  {title} {\bibinfo {title} {High-throughput all-atom molecular dynamics
  simulations using distributed computing},\ }\href@noop {} {\bibfield
  {journal} {\bibinfo  {journal} {J. Chem. Theory Comput.}\ }\textbf {\bibinfo
  {volume} {50}},\ \bibinfo {pages} {397} (\bibinfo {year} {2010})}\BibitemShut
  {NoStop}%
\bibitem [{\citenamefont {Ciccotti}\ \emph {et~al.}(2008)\citenamefont
  {Ciccotti}, \citenamefont {Lelievre},\ and\ \citenamefont
  {Vanden-Eijnden}}]{ciccotti2008projection}%
  \BibitemOpen
  \bibfield  {author} {\bibinfo {author} {\bibfnamefont {G.}~\bibnamefont
  {Ciccotti}}, \bibinfo {author} {\bibfnamefont {T.}~\bibnamefont {Lelievre}},\
  and\ \bibinfo {author} {\bibfnamefont {E.}~\bibnamefont {Vanden-Eijnden}},\
  }\bibfield  {title} {\bibinfo {title} {Projection of diffusions on
  submanifolds: Application to mean force computation},\ }\href@noop {}
  {\bibfield  {journal} {\bibinfo  {journal} {Commun. Pure Appl. Math.}\
  }\textbf {\bibinfo {volume} {61}},\ \bibinfo {pages} {371} (\bibinfo {year}
  {2008})}\BibitemShut {NoStop}%
\bibitem [{\citenamefont {Kalligiannaki}\ \emph {et~al.}(2015)\citenamefont
  {Kalligiannaki}, \citenamefont {Harmandaris}, \citenamefont {Katsoulakis},\
  and\ \citenamefont {Plech{\'a}{\v{c}}}}]{kalligiannaki2015geometry}%
  \BibitemOpen
  \bibfield  {author} {\bibinfo {author} {\bibfnamefont {E.}~\bibnamefont
  {Kalligiannaki}}, \bibinfo {author} {\bibfnamefont {V.}~\bibnamefont
  {Harmandaris}}, \bibinfo {author} {\bibfnamefont {M.~A.}\ \bibnamefont
  {Katsoulakis}},\ and\ \bibinfo {author} {\bibfnamefont {P.}~\bibnamefont
  {Plech{\'a}{\v{c}}}},\ }\bibfield  {title} {\bibinfo {title} {The geometry of
  generalized force matching and related information metrics in coarse-graining
  of molecular systems},\ }\href@noop {} {\bibfield  {journal} {\bibinfo
  {journal} {J. Chem. Phys.}\ }\textbf {\bibinfo {volume} {143}} (\bibinfo
  {year} {2015})}\BibitemShut {NoStop}%
\bibitem [{\citenamefont {Torrie}\ and\ \citenamefont
  {Valleau}(1977)}]{torrie1977nonphysical}%
  \BibitemOpen
  \bibfield  {author} {\bibinfo {author} {\bibfnamefont {G.~M.}\ \bibnamefont
  {Torrie}}\ and\ \bibinfo {author} {\bibfnamefont {J.~P.}\ \bibnamefont
  {Valleau}},\ }\bibfield  {title} {\bibinfo {title} {Nonphysical sampling
  distributions in monte carlo free-energy estimation: Umbrella sampling},\
  }\href@noop {} {\bibfield  {journal} {\bibinfo  {journal} {J. Comput. Phys.}\
  }\textbf {\bibinfo {volume} {23}},\ \bibinfo {pages} {187} (\bibinfo {year}
  {1977})}\BibitemShut {NoStop}%
\bibitem [{\citenamefont {Rosso}\ \emph {et~al.}(2002)\citenamefont {Rosso},
  \citenamefont {Min{\'a}ry}, \citenamefont {Zhu},\ and\ \citenamefont
  {Tuckerman}}]{rosso2002use}%
  \BibitemOpen
  \bibfield  {author} {\bibinfo {author} {\bibfnamefont {L.}~\bibnamefont
  {Rosso}}, \bibinfo {author} {\bibfnamefont {P.}~\bibnamefont {Min{\'a}ry}},
  \bibinfo {author} {\bibfnamefont {Z.}~\bibnamefont {Zhu}},\ and\ \bibinfo
  {author} {\bibfnamefont {M.~E.}\ \bibnamefont {Tuckerman}},\ }\bibfield
  {title} {\bibinfo {title} {On the use of the adiabatic molecular dynamics
  technique in the calculation of free energy profiles},\ }\href@noop {}
  {\bibfield  {journal} {\bibinfo  {journal} {J. Chem. Phys.}\ }\textbf
  {\bibinfo {volume} {116}},\ \bibinfo {pages} {4389} (\bibinfo {year}
  {2002})}\BibitemShut {NoStop}%
\bibitem [{\citenamefont {Maragliano}\ and\ \citenamefont
  {Vanden-Eijnden}(2006)}]{maragliano2006temperature}%
  \BibitemOpen
  \bibfield  {author} {\bibinfo {author} {\bibfnamefont {L.}~\bibnamefont
  {Maragliano}}\ and\ \bibinfo {author} {\bibfnamefont {E.}~\bibnamefont
  {Vanden-Eijnden}},\ }\bibfield  {title} {\bibinfo {title} {A temperature
  accelerated method for sampling free energy and determining reaction pathways
  in rare events simulations},\ }\href@noop {} {\bibfield  {journal} {\bibinfo
  {journal} {Chem. Phys. Lett.}\ }\textbf {\bibinfo {volume} {426}},\ \bibinfo
  {pages} {168} (\bibinfo {year} {2006})}\BibitemShut {NoStop}%
\bibitem [{\citenamefont {Hastie}\ \emph {et~al.}(2009)\citenamefont {Hastie},
  \citenamefont {Tibshirani}, \citenamefont {Friedman},\ and\ \citenamefont
  {Friedman}}]{hastie2009elements}%
  \BibitemOpen
  \bibfield  {author} {\bibinfo {author} {\bibfnamefont {T.}~\bibnamefont
  {Hastie}}, \bibinfo {author} {\bibfnamefont {R.}~\bibnamefont {Tibshirani}},
  \bibinfo {author} {\bibfnamefont {J.~H.}\ \bibnamefont {Friedman}},\ and\
  \bibinfo {author} {\bibfnamefont {J.~H.}\ \bibnamefont {Friedman}},\
  }\href@noop {} {\emph {\bibinfo {title} {The elements of statistical
  learning: data mining, inference, and prediction}}},\ Vol.~\bibinfo {volume}
  {2}\ (\bibinfo  {publisher} {Springer},\ \bibinfo {year} {2009})\BibitemShut
  {NoStop}%
\bibitem [{\citenamefont {Bishop}\ and\ \citenamefont
  {Nasrabadi}(2006)}]{bishop2006pattern}%
  \BibitemOpen
  \bibfield  {author} {\bibinfo {author} {\bibfnamefont {C.~M.}\ \bibnamefont
  {Bishop}}\ and\ \bibinfo {author} {\bibfnamefont {N.~M.}\ \bibnamefont
  {Nasrabadi}},\ }\href@noop {} {\emph {\bibinfo {title} {Pattern recognition
  and machine learning}}},\ Vol.~\bibinfo {volume} {4}\ (\bibinfo  {publisher}
  {Springer},\ \bibinfo {year} {2006})\BibitemShut {NoStop}%
\bibitem [{\citenamefont {Noid}\ \emph {et~al.}(2007)\citenamefont {Noid},
  \citenamefont {Chu}, \citenamefont {Ayton},\ and\ \citenamefont
  {Voth}}]{noid2007multiscale}%
  \BibitemOpen
  \bibfield  {author} {\bibinfo {author} {\bibfnamefont {W.}~\bibnamefont
  {Noid}}, \bibinfo {author} {\bibfnamefont {J.-W.}\ \bibnamefont {Chu}},
  \bibinfo {author} {\bibfnamefont {G.~S.}\ \bibnamefont {Ayton}},\ and\
  \bibinfo {author} {\bibfnamefont {G.~A.}\ \bibnamefont {Voth}},\ }\bibfield
  {title} {\bibinfo {title} {Multiscale coarse-graining and structural
  correlations: Connections to liquid-state theory},\ }\href@noop {} {\bibfield
   {journal} {\bibinfo  {journal} {J. Phys. Chem. B}\ }\textbf {\bibinfo
  {volume} {111}},\ \bibinfo {pages} {4116} (\bibinfo {year}
  {2007})}\BibitemShut {NoStop}%
\bibitem [{\citenamefont {Shell}(2008)}]{shell2008relative}%
  \BibitemOpen
  \bibfield  {author} {\bibinfo {author} {\bibfnamefont {M.~S.}\ \bibnamefont
  {Shell}},\ }\bibfield  {title} {\bibinfo {title} {The relative entropy is
  fundamental to multiscale and inverse thermodynamic problems},\ }\href
  {https://doi.org/10.1063/1.2992060} {\bibfield  {journal} {\bibinfo
  {journal} {J. Chem. Phys.}\ }\textbf {\bibinfo {volume} {129}},\ \bibinfo
  {pages} {144108} (\bibinfo {year} {2008})}\BibitemShut {NoStop}%
\bibitem [{\citenamefont {Mullinax}\ and\ \citenamefont
  {Noid}(2009)}]{mullinax2009generalized}%
  \BibitemOpen
  \bibfield  {author} {\bibinfo {author} {\bibfnamefont {J.}~\bibnamefont
  {Mullinax}}\ and\ \bibinfo {author} {\bibfnamefont {W.}~\bibnamefont
  {Noid}},\ }\bibfield  {title} {\bibinfo {title} {Generalized yvon-born-green
  theory for molecular systems},\ }\href@noop {} {\bibfield  {journal}
  {\bibinfo  {journal} {Phys. Rev. Lett.}\ }\textbf {\bibinfo {volume} {103}},\
  \bibinfo {pages} {198104} (\bibinfo {year} {2009})}\BibitemShut {NoStop}%
\bibitem [{\citenamefont {Rudzinski}\ and\ \citenamefont
  {Noid}(2011)}]{rudzinski2011coarse}%
  \BibitemOpen
  \bibfield  {author} {\bibinfo {author} {\bibfnamefont {J.~F.}\ \bibnamefont
  {Rudzinski}}\ and\ \bibinfo {author} {\bibfnamefont {W.}~\bibnamefont
  {Noid}},\ }\bibfield  {title} {\bibinfo {title} {Coarse-graining entropy,
  forces, and structures},\ }\href@noop {} {\bibfield  {journal} {\bibinfo
  {journal} {J. Chem. Phys.}\ }\textbf {\bibinfo {volume} {135}} (\bibinfo
  {year} {2011})}\BibitemShut {NoStop}%
\bibitem [{\citenamefont {Fu}\ \emph {et~al.}(2022)\citenamefont {Fu},
  \citenamefont {Wu}, \citenamefont {Wang}, \citenamefont {Xie}, \citenamefont
  {Keten}, \citenamefont {Gomez-Bombarelli},\ and\ \citenamefont
  {Jaakkola}}]{fu2022forces}%
  \BibitemOpen
  \bibfield  {author} {\bibinfo {author} {\bibfnamefont {X.}~\bibnamefont
  {Fu}}, \bibinfo {author} {\bibfnamefont {Z.}~\bibnamefont {Wu}}, \bibinfo
  {author} {\bibfnamefont {W.}~\bibnamefont {Wang}}, \bibinfo {author}
  {\bibfnamefont {T.}~\bibnamefont {Xie}}, \bibinfo {author} {\bibfnamefont
  {S.}~\bibnamefont {Keten}}, \bibinfo {author} {\bibfnamefont
  {R.}~\bibnamefont {Gomez-Bombarelli}},\ and\ \bibinfo {author} {\bibfnamefont
  {T.}~\bibnamefont {Jaakkola}},\ }\bibfield  {title} {\bibinfo {title} {Forces
  are not enough: Benchmark and critical evaluation for machine learning force
  fields with molecular simulations},\ }\href@noop {} {\bibfield  {journal}
  {\bibinfo  {journal} {arXiv preprint arXiv:2210.07237}\ } (\bibinfo {year}
  {2022})}\BibitemShut {NoStop}%
\bibitem [{\citenamefont {Duschatko}\ \emph
  {et~al.}(2024{\natexlab{b}})\citenamefont {Duschatko}, \citenamefont
  {Vandermause}, \citenamefont {Molinari},\ and\ \citenamefont
  {Kozinsky}}]{duschatko2024uncertainty}%
  \BibitemOpen
  \bibfield  {author} {\bibinfo {author} {\bibfnamefont {B.~R.}\ \bibnamefont
  {Duschatko}}, \bibinfo {author} {\bibfnamefont {J.}~\bibnamefont
  {Vandermause}}, \bibinfo {author} {\bibfnamefont {N.}~\bibnamefont
  {Molinari}},\ and\ \bibinfo {author} {\bibfnamefont {B.}~\bibnamefont
  {Kozinsky}},\ }\bibfield  {title} {\bibinfo {title} {Uncertainty driven
  active learning of coarse grained free energy models},\ }\href@noop {}
  {\bibfield  {journal} {\bibinfo  {journal} {npj Comput. Mater.}\ }\textbf
  {\bibinfo {volume} {10}},\ \bibinfo {pages} {9} (\bibinfo {year}
  {2024}{\natexlab{b}})}\BibitemShut {NoStop}%
\end{thebibliography}%

\end{document}